\documentclass[11pt]{article}

\usepackage[margin=1in]{geometry}
\usepackage{amsmath,amsthm,amssymb,mathtools,bm}
\usepackage{bbm}
\usepackage{enumitem}
\usepackage{hyperref}
\hypersetup{colorlinks=true,allcolors=black,citecolor=blue}%
\usepackage{xr}
\usepackage{comment}
\usepackage{xcolor}
\definecolor{brown}{RGB}{139,69,19}
\definecolor{purple}{RGB}{128,0,128}
\usepackage{amsbsy}
\usepackage{pdflscape}
\usepackage{subcaption}
\usepackage{float}
\usepackage{dsfont}
\usepackage{booktabs}
\usepackage{longtable}
\usepackage{multirow}
\usepackage{graphicx}
\usepackage{setspace}
\usepackage{algorithm}
\usepackage{placeins}
\usepackage{needspace}
\usepackage{natbib}
\usepackage{authblk}

\newcommand{\algorithmspacing}{%
  \raggedright\setstretch{1.02}\setdisplayskipstretch{1}%
  \setlength{\parskip}{2pt}%
  \setlength{\abovedisplayskip}{4pt}%
  \setlength{\belowdisplayskip}{4pt}%
  \setlength{\abovedisplayshortskip}{3pt}%
  \setlength{\belowdisplayshortskip}{3pt}%
}%
\newtheorem{assumption}{Assumption}
\newtheorem{theorem}{Theorem}
\newtheorem{lemma}{Lemma}
\newtheorem{proposition}{Proposition}
\newtheorem{definition}{Definition}
\newtheorem{corollary}{Corollary}
\theoremstyle{remark}
\newtheorem{remark}{Remark}
\newtheorem{example}{Example}

\newcommand{\E}{\mathbb{E}}

\newcommand{\1}{\mathbbm{1}}

\graphicspath{{_support/figure/}}
\title{Conformal Prediction for \textcolor{black}{Multi-Step} Forecasting:\\
\textcolor{black}{Conditional Coverage and Sharp Score Calibration Rates}}

\author[1]{Siyuan He}
\author[2,3]{Yuhong Yang\thanks{\textcolor{black}{Corresponding author: \texttt{yyangsc@mail.tsinghua.edu.cn}.}}}
\affil[1]{Qiuzhen College, Tsinghua University, Beijing, China}
\affil[2]{Yau Mathematical Sciences Center, Tsinghua University, Beijing, China}
\affil[3]{Beijing Institute of Mathematical Sciences and Applications, Beijing, China}
\date{\textcolor{black}{\today}}

\begin{document}
\setstretch{1.2}
\raggedbottom
\maketitle
\begin{abstract}
{\color{black}\hypersetup{allcolors=black,citecolor=blue}%
Uncertainty quantification in prediction is a central problem in statistics
and machine learning. {\color{black}Conformal prediction provides a flexible approach to constructing prediction
sets with valid marginal coverage under exchangeability.}
In this paper, we investigate conformal prediction for both one-step and
\textcolor{black}{multi-step} forecasting from a single stationary time series, addressing
temporal dependence and the additional challenges of joint prediction
across multiple horizons. We propose a kernel-based \textcolor{black}{framework, called \textcolor{black}{Time series} Residual Analysis for Calibration with Kernels (TRACK),} for estimating conditional
quantiles of residual scores and constructing prediction sets from
rolling-window forecasters. The framework accommodates heterogeneous
uncertainty across horizons and accounts for dependence induced by rolling
estimation, while allowing predictor memory and forecast horizons to grow
with the sample size. {\color{black}Our finite-sample bounds quantify how temporal dependence and overlapping
score windows affect calibration and coverage. \textcolor{black}{We further derive
bounds for conditional quantile estimation and pointwise
conditional \textcolor{black}{coverage discrepancy}.}} {\color{black}\hypersetup{allcolors=black,citecolor=blue}The guarantees extend to more general predictors that admit an
accurate rolling-window approximation.} We also derive matching
minimax bounds over classes of known score maps, clarifying how overlapping
score windows affect calibration accuracy. {\color{black}Experimental results from
simulations and applications to {\color{black}wind energy generation and \textcolor{black}{bike sharing} demand}
show that our methods improve regional conditional coverage while
maintaining competitive interval widths.}
}
\end{abstract}

{\color{black}\hypersetup{allcolors=black,citecolor=blue}%
\par}

\newpage
\section{Introduction}
In modern statistics and machine learning, uncertainty quantification is
of fundamental importance for predictive analysis. While point predictions are often
useful summaries, they provide only limited information about the
uncertainty surrounding future outcomes and can therefore be
inadequate for downstream \textcolor{black}{decision making}. This limitation is
especially pronounced when predictive models are highly nonlinear,
data-adaptive \textcolor{black}{or} otherwise difficult to analyze theoretically.
Conformal prediction {\color{blue}\citep{vovk2005algorithmic}} offers a flexible
framework for addressing this problem. Under exchangeability, it
constructs prediction sets with finite-sample marginal coverage
guarantees {\color{blue}\citep{shafer2008tutorial,lei2014distribution}}, without
requiring a correctly specified predictive model
{\color{black}\hypersetup{allcolors=black,citecolor=blue}{\color{blue}\citep{lei2018distribution}}}.
{\color{black}\hypersetup{allcolors=black,citecolor=blue}Because of its distribution-free validity, conformal prediction is particularly
attractive for modern applications involving complex or black-box models.}

The extension of conformal prediction beyond exchangeability has
attracted increasing attention. For time series,
serial dependence changes the relationship between past conformity
scores and future prediction errors.
{\color{black}\hypersetup{allcolors=black,citecolor=blue}Existing research
\textcolor{black}{develops sequential and adaptive calibration methods}
{\color{blue}(\citealp{gibbs2021adaptive,
xu2023sequential,gibbs2024conformal}{\color{blue}\hypersetup{citecolor=blue};
\citealp{humbert2026online}})}, while recent theoretical work
studies split conformal prediction under nonexchangeability
{\color{blue}\citep{oliveira2024split}} and derives marginal coverage bounds allowing
predictor memory to grow
{\color{blue}\citep{barber2025predictive}}.
Weighted empirical quantiles have also been used to calibrate prediction
intervals
{\color{blue}\citep{barber2023conformal,lee2025kernel}}.}

{\color{black}\textcolor{black}{Multi-step}-ahead forecasting constitutes an important component of \textcolor{black}{time series} analysis.} In applications such as temperature
forecasting {\color{blue}\citep{fang2021multi}} and macroeconomic policy analysis
{\color{blue}\citep{clark2025constructing}}, decisions depend on forecasts over
several future periods. Forecast errors at different horizons are
dependent and can have substantially different scales. {\color{black}Moreover, uncertainty quantification for the full future trajectory \textcolor{black}{up to a certain time ahead} calls for
joint coverage guarantees across forecast horizons.}
{\color{black}\hypersetup{allcolors=black,citecolor=blue}These challenges have motivated a growing literature on
conformal prediction for \textcolor{black}{multihorizon} and multivariate outcomes
{\color{blue}\citep{stankeviciute2021conformal,sun2023copula,
english2025janet}}.}
{\color{black}\hypersetup{allcolors=black,citecolor=blue}%
Adaptive forecasting bands have been developed for heterogeneous,
exchangeable trajectories {\color{blue}\citep{zhou2024conformalized}}.
For multivariate outcomes, optimal-transport ranks have been used to
construct conformal prediction regions under exchangeability
{\color{blue}\citep{thurin2025optimal}}.}

{\color{black}\hypersetup{allcolors=black,citecolor=blue}In this paper, we develop a unified framework \textcolor{black}{called \textcolor{black}{Time series} Residual Analysis for Calibration with Kernels (TRACK)} for one-step
and \textcolor{black}{multi-step} prediction from a single \textcolor{black}{stationary} time series using
general rolling-window forecasters. A normalized residual score
accommodates heterogeneous uncertainty across forecast horizons and
reduces joint prediction to conditional quantile estimation.
Kernel-based calibration then yields prediction sets that adapt to
\textcolor{black}{the new covariate value}. Our analysis accounts for the dependence
induced by rolling estimation and allows predictor memory to grow
with the sample size.} {\color{black}\hypersetup{allcolors=black,citecolor=blue}%
We further relax this restriction to include predictors
that admit an accurate rolling-window approximation.}
{\color{black}\hypersetup{allcolors=black,citecolor=blue}%
To our knowledge, this is the first work to establish matching minimax bounds
for \textcolor{black}{score calibration} \textcolor{black}{over scoring rules with a given window span
in a single time series, while allowing predictor memory and forecast horizons to grow.} \textcolor{black}{For the proposed TRACK approach, this work makes three theoretical
contributions.}

First, we establish finite-sample bounds for calibration error and the
resulting marginal and group-conditional \textcolor{black}{coverage discrepancies}, covering both
independent benchmark targets and future targets from the same series.
These bounds account for temporal dependence and the additional dependence
induced by rolling estimation and overlapping score windows. They quantify
how predictor memory and forecast horizons affect calibration accuracy.

\textcolor{black}{Second, we derive bounds for conditional quantile estimation
and pointwise conditional \textcolor{black}{coverage discrepancy}. Under suitable
regularity conditions, these bounds apply to both one-step intervals and joint
multi-step prediction regions for future observations from the same series,
allowing predictor memory and forecast horizons to grow with sample size.}

Third, we derive minimax rates over \textcolor{black}{given} score maps with a
prescribed window span and show that the span alone does not determine the
calibration difficulty of a specified score map. The matching bounds identify
the best calibration accuracy attainable uniformly over this class. In
particular, scoring rules with the same span can have different optimal
rates, showing that how historical information enters the score can
fundamentally affect calibration difficulty.
\par%
}%

{\color{black}\hypersetup{allcolors=black,citecolor=blue}%
We evaluate \textcolor{black}{\textcolor{black}{TRACK} and its variants} through simulations and applications to
hourly generation at the Hackberry Wind Project {\color{blue}\citep{xu2023conformal}}
and hourly demand in the Capital Bikeshare system
{\color{blue}\citep{fanaee2014event,romano2019conformalized}}.
{\color{black}Overall, our methods achieve lower conditional coverage \textcolor{black}{deviations} and perform
competitively in marginal coverage and interval width.}}

\Needspace{7\baselineskip}
The rest of the paper is organized as follows.
{\color{black}\hypersetup{allcolors=black,citecolor=blue}%
Section~{\color{black}\ref{sec:setup}} formulates the rolling forecasting problem and defines
the marginal and conditional coverage targets.
Section~{\color{black}\ref{sec:mth}} \textcolor{black}{presents \textcolor{black}{TRACK} and its variants, which construct} one-step intervals and joint \textcolor{black}{multi-step}
prediction regions through local calibration of forecast errors.
Section~{\color{black}\ref{sec:upper-bounds}} quantifies finite-sample calibration and coverage
\textcolor{black}{discrepancy} under temporal dependence and overlapping score windows.
{\color{black}\hypersetup{allcolors=black,citecolor=blue}Section~\ref{sec:conditional-validity}
gives bounds for conditional quantile estimation and conditional \textcolor{black}{coverage discrepancy}.}
Section~{\color{black}\ref{sec:minimax}} derives matching minimax bounds and examines how
\textcolor{black}{score window} structure affects calibration difficulty.
Section~{\color{black}\ref{sec:num}} evaluates coverage and \textcolor{black}{prediction set} size through
simulations and real data applications, and
\color{black}\hypersetup{allcolors=black,citecolor=blue}%
Section~{\color{black}\ref{sec:disc}} concludes with a discussion. {\color{black}\hypersetup{allcolors=black,citecolor=blue}Technical proofs and additional discussions are provided
in the supplementary material.}}

\paragraph{Notation.}
{\color{black}\hypersetup{allcolors=black,citecolor=blue}%
We write $\mathbb N_0=\mathbb N\cup\{0\}$ and
$[m]=\{1,\ldots,m\}$ for $m\in\mathbb N$.
The norm $\|\cdot\|$ is Euclidean, and $\|f\|_\infty=\sup_x|f(x)|$.
The ball centered at $x$ with radius $r$ is
$B(x,r)=\{y:\|y-x\|<r\}$.
For real numbers $a,b$, we use $a\wedge b=\min(a,b)$ and
$a\vee b=\max(a,b)$. The cardinality of a finite set $A$ is $|A|$,
and $A\mathbin\triangle B$ denotes symmetric difference.
Indicators and Lebesgue measure are written as $\1\{\cdot\}$ and
$\operatorname{Leb}$ respectively, while $\mathcal A\vee\mathcal B$
is the sigma-field generated by $\mathcal A$ and $\mathcal B$.
The gamma function \textcolor{black}{is defined by} $\Gamma(z)=\int_0^\infty t^{z-1}e^{-t}\,dt$, $z>0$.
For a law $P$, $P_X$ is its covariate marginal, while $P\{\cdot\}$
and $\mathbb E_P$ denote probabilities and expectations, written
$\mathbb P$ and $\mathbb E$ when the law is fixed.
{\color{black}\hypersetup{allcolors=black,citecolor=blue}Randomized rules use an internal seed $V$ independent of the entire
data-generating process.}
Unless conditioning is specified, probabilities and expectations average
over the data, target \textcolor{black}{and} algorithmic randomness.
For positive sequences, $u_N\asymp v_N$ means that their ratio is bounded
above and below by positive constants independent of $N$, and
$u_N=o(v_N)$ means $u_N/v_N\to0$.
We write $\xrightarrow{P}$ for convergence in probability.
Generic positive constants $c,C$ may change from line to line.}
\begingroup%
\color{black}%
\hypersetup{allcolors=black,citecolor=blue}%
\par
\section[Forecasting setup and coverage targets]{\textcolor{black}{Forecasting setup and coverage targets}}
\label{sec:setup}
Let $\{Z_t=(X_t,Y_t)\}_{t\in{\mathbb N}}$ be a strictly stationary time series{\color{black}\hypersetup{allcolors=black,citecolor=blue} taking values in $\mathcal Z=\mathbb R^d\times\mathbb R$}, where $X_t\in\mathcal X\subseteq\mathbb R^d$ denotes the covariate vector and $Y_t\in\mathcal Y\subseteq\mathbb R$ denotes the corresponding response. {{Suppose that $N$ data pairs $(Z_1,\ldots,Z_N)$ are observed together
with the covariate $X_{N+1}$.} In the asymptotic analysis, $N\to\infty$ means \textcolor{black}{increasingly longer series}
of the same strictly stationary process under a fixed law.}

{{At each {\color{black}\hypersetup{allcolors=black,citecolor=blue}forecast origin} $t$, we use the observed history through time $t$ and the covariate $X_{t+1}$ to quantify uncertainty about future responses.}} {{For a deterministic horizon $H=H_N\in\mathbb N$, {let $\{g_h\}_{h=1}^H$ be a collection of point \textcolor{black}{forecasts},
where $g_h=g_{N,h}$ is designed to predict the $h$-step-ahead response.} {\color{black}\hypersetup{allcolors=black,citecolor=blue}The forecasting algorithms may be randomized,
with internal randomness independent of the entire data-generating process.}}} {\color{black}\hypersetup{allcolors=black,citecolor=blue}In time series applications, forecasting models are often updated sequentially as new observations become available, and it is common in practice to use only the most recent data to obtain a local approximation to the predictive relationship.} {To accommodate such scenarios,} \textcolor{black}{choose} {$L=L_N\in\mathbb N_0$} and allow each predictor to be trained using a rolling window of the most recent $L$ observations. {\textcolor{black}{For each $N$}, the same forecasting rules are applied across origins
\textcolor{black}{$t\in\{L_N,\ldots,N\}$}, giving}
{\color{black}%
\begin{equation}\label{eq:point_forecast}
    \widehat Y_{N,t+h\mid t}
    =g_{N,h}\!\left(X_{t+1};\,Z_t,Z_{t-1},\ldots,Z_{t-L_N+1}\right),
    \qquad h\in[H_N],\quad \textcolor{black}{t\in\{L_N,\ldots,N\}}.
\end{equation}}%
{Here, ${\color{black}\widehat Y_{N,t+h\mid t}}$ denotes the prediction of $Y_{t+h}$
formed at forecast origin $t$.} This formulation allows the forecasting rule to depend on recent observations in a flexible way. \textcolor{black}{When $L_N=0$, the history argument is empty. This covers forecasts from pretrained models, such as those trained offline on an external dataset.}
Therefore, the above framework provides a formulation for both adaptive rolling-window predictors and externally trained forecasting models, which will serve as the basis for our subsequent construction of prediction sets.
{\color{black}\hypersetup{allcolors=black,citecolor=blue}Sections~\ref{subsec:general-predictors} and~\ref{subsec:general-conditional-validity} allow predictors fitted to longer
histories, including all available observations, when they are close to
rolling-window predictors.}

\Needspace{3\baselineskip}
{{\color{black}\hypersetup{allcolors=black,citecolor=blue}Given the information available at time $t=N$ and the point \textcolor{black}{forecasts} $\{g_h\}_{h=1}^H$}}, our objective is to construct a data-dependent prediction set
$\widehat C_{N}(X_{{N}+1}) \subseteq \mathbb R^H$
{for the future response vector
$Y_{{N}+1:{N}+H}:=(Y_{{N}+1},\ldots,Y_{{N}+H})^\top\in\mathbb R^H$.}
In other words, rather than quantifying uncertainty separately for each forecasting horizon, we aim to provide a joint prediction region for the entire future trajectory over the next $H$ time points.
{\color{black}\hypersetup{allcolors=black,citecolor=blue}Such a formulation is particularly relevant for \textcolor{black}{multi-step}-ahead
forecasting, where decisions may depend on the joint behavior of several
future responses {\color{blue}\citep{wolf2015bootstrap}}.}

Fix a miscoverage level $\alpha\in(0,1)$. Ideally, one would like the prediction set $\widehat C_{N}(\cdot)$ to satisfy the {finite-sample marginal coverage guarantee}
\begin{equation*}\color{black}
	\mathbb P\!\left\{Y_{{N}+1:{N}+H}\in \widehat C_{N}(X_{{N}+1})\right\}=1-\alpha,
\end{equation*}%
which requires the prediction set to contain the future response vector with probability $1-\alpha$ on average over the joint distribution of the data. While marginal coverage is a standard guarantee in conformal prediction, it does not describe how the coverage probability may vary across different values of the covariate $X_{{N}+1}$. In applications where the uncertainty of future responses depends strongly on the covariates, a stronger and more informative target is the {finite-sample conditional coverage guarantee}
\begin{equation}\label{eq:conditional_ideal}
	\mathbb P\!\left\{Y_{{N}+1:{N}+H}\in \widehat C_{N}(X_{{N}+1}) \,\big|\, X_{{N}+1}=x\right\}=1-\alpha,
\end{equation}%
{for $P_X$-almost every $x\in\mathcal X$.}

{{\color{black}\hypersetup{allcolors=black,citecolor=blue}However, \textcolor{black}{as is well known,} requiring distribution-free finite-sample conditional coverage
of at least $1-\alpha$ can force uninformative prediction sets
{\color{blue}({\color{blue}\hypersetup{citecolor=blue}\citealp{pmlr-v25-vovk12}}; \citealp{lei2014distribution})}.}} For this reason, it is natural to seek weaker but still meaningful notions of conditional validity that remain both theoretically tractable and practically useful.

To motivate such a relaxation, observe that the ideal conditional coverage
requirement {\color{black}\eqref{eq:conditional_ideal}} is equivalent to
\begin{equation}\label{eq:between_ideal}
	\mathbb E\!\left[
		f(X_{{N}+1})
		\Bigl\{
		\1\{Y_{{N}+1:{N}+H}\in\widehat C_{N}(X_{{N}+1})\}
		-(1-\alpha)
		\Bigr\}
	\right]=0
\end{equation}%
for every bounded measurable function $f$.
{{Restricting the test functions in {\color{black}\eqref{eq:between_ideal}} to a prescribed
class $\mathcal F$ gives weaker coverage requirements {\color{blue}\citep{gibbs2025conformal}}.
The choice $\mathcal F=\{1\}$ yields marginal coverage, while group
indicators yield coverage within the corresponding groups. {\color{black}\hypersetup{allcolors=black,citecolor=blue}Allowing a tolerance ${\color{black}\hypersetup{allcolors=black,citecolor=blue}\eta_N}\ge0$ for deviations from exact
coverage, we consider the following approximate conditional coverage
requirement:}}}
{%
\begin{equation}\label{eq:between}
\sup_{\|f\|_\infty\le1}
\left|\mathbb E\!\left[
f(X_{{N}+1})\Bigl\{
\1\{Y_{{N}+1:{N}+H}\in\widehat C_{N}(X_{{N}+1})\}-(1-\alpha)
\Bigr\}\right]\right|
\le{\color{black}\hypersetup{allcolors=black,citecolor=blue}\eta_N}.
\end{equation}}%

\Needspace{5\baselineskip}
{{{\color{black}\hypersetup{allcolors=black,citecolor=blue}{\color{black}Our goal is to construct informative prediction sets that adapt to
covariate-dependent uncertainty while accounting for temporal dependence.
For \textcolor{black}{multi-step} forecasting, validity must concern the entire future
trajectory \textcolor{black}{up to time $N+H_N$}, and the resulting regions should accommodate differences in
uncertainty across horizons as well as dependence among future responses.
We therefore seek a unified approach to one-step and \textcolor{black}{multi-step} prediction
that supports conditional validity while allowing the forecast horizon to
increase with the available sample size.}}}}

\endgroup%

\par
\color{black}%
\section[Local calibration for time series forecasting]{\textcolor{black}{Local calibration for \textcolor{black}{time series} forecasting}}
\label{sec:mth}
\color{black}%
\begingroup%
\color{black}%
\hypersetup{allcolors=black,citecolor=blue}%
\setlength{\abovedisplayskip}{8pt plus 1pt minus 2pt}
\setlength{\belowdisplayskip}{8pt plus 1pt minus 2pt}
\setlength{\abovedisplayshortskip}{4pt plus 1pt}
\setlength{\belowdisplayshortskip}{6pt plus 1pt minus 2pt}
\subsection{One-step-ahead prediction}\label{sec:one_step}

\begin{algorithm}[!t]
	\caption{\textcolor{black}{\textcolor{black}{TRACK}:} One-Step-Ahead Prediction Interval}
	\label{alg:quantile_estimation}
	\algorithmspacing
	\textbf{Input:} Data $\{Z_1,\ldots,Z_{N}, X_{{N}+1}\}$, predictor ${g=g_{N,1}}$, kernel $K$, \mbox{training length ${L_N}$}, \mbox{bandwidth $h_{N}$}, significance level $\alpha$.\par
	\textbf{Output:} Conditional quantile estimators ${\color{black}{\color{black}\widehat q^{\mathrm{ker}}_{N,\alpha/2}}}(x)$ and ${\color{black}{\color{black}\widehat q^{\mathrm{ker}}_{N,1-\alpha/2}}}(x)$, \mbox{prediction interval $\widehat{C}_{N}(X_{{N}+1})$}.\par
	\begin{enumerate}[label=\arabic*.,leftmargin=1.6em,labelsep=0.55em,topsep=5pt,itemsep=5pt,parsep=0pt,partopsep=0pt]
		\item \textcolor{black}{For $t=L_N+1,\ldots,N+1$, compute} the estimator of $Y_t$ based on predictor $g$ and the past ${L_N}$ observations.
		\[
{%
        \widehat Y_{N,t}=g\bigl(X_t;Z_{t-1},\ldots,Z_{t-L_N}\bigr).
        }%
\]%
		\item Define the residuals ${R_{N,t}:=Y_t-\widehat Y_{N,t}}$ and calculate the locally weighted empirical \mbox{distribution function}
		\[
{%
		{\widehat F_N}(u;x):=\frac{\sum_{t={L_N}+1}^{N} K_{h_{N}}(X_t-x)\,\1\{R_{N,t}\le u\}}{\sum_{t={L_N}+1}^{N} K_{h_{N}}(X_t-x)}.
		}%
\]%
		\item For each $x \in \mathcal X$ and ${\tau\in(0,1)}$, \textcolor{black}{define}
		\[
		{\color{black}{\color{black}\widehat q^{\mathrm{ker}}_{N,\tau}}}(x):=\inf\{u:{\widehat F_N}(u;x)\ge\tau\}.
		\]%
		\item The prediction interval is \textcolor{black}{obtained as}
		\[
{%
		\widehat{C}_{N}(X_{{N}+1})=\Big[\widehat Y_{N,N+1}+{\color{black}{\color{black}\widehat q^{\mathrm{ker}}_{N,\alpha/2}}}(X_{{N}+1}), \ \widehat Y_{N,N+1}+{\color{black}{\color{black}\widehat q^{\mathrm{ker}}_{N,1-\alpha/2}}}(X_{{N}+1})\Big].
		}%
\]%
	\end{enumerate}
\end{algorithm}

{\color{black}We begin with one-step-ahead \textcolor{black}{prediction}. Our objective is
to estimate \textcolor{black}{forecast error} \textcolor{black}{distributions} conditional on the covariates.}
{Let $g:=g_{N,1}$ denote the one-step predictor. At time $t$, its
forecast based on the most recent $L_N$ observations is
$\widehat Y_{N,t}=g(X_t;Z_{t-1},\ldots,Z_{t-L_N})$.
{\color{black}The corresponding forecast error is then defined as $R_{N,t}:=Y_t-\widehat Y_{N,t}$.}}
{\color{black}For $\tau\in(0,1)$, we define the conditional cumulative distribution
function and corresponding quantile by}
\[
{%
F_N(u\mid x):=\mathbb P\{R_{N,t}\le u\mid X_t=x\},
\qquad
q_{N,\tau}(x):=\inf\{u:F_N(u\mid x)\ge\tau\}.}
\]%
{For each fixed $N$, the forecasting rule is time invariant and uses
the same window length $L_N$ across origins. Strict stationarity of
$\{Z_t\}$ therefore implies strict stationarity of $\{(X_t,R_{N,t})\}$, so
$F_N(\cdot\mid x)$ does not depend on $t$.}

{\color{black}\hypersetup{allcolors=black,citecolor=blue}%
Conditional distribution estimation methods, such as histogram-based
approaches, have been used to construct conformal prediction intervals
{\color{blue}\citep{sesia2021conformal}}.}
{\textcolor{black}{In \textcolor{black}{TRACK}, we} estimate $F_N(\cdot\mid x)$ by a kernel-weighted empirical
{distribution function} of the calibration residuals and
estimate its quantiles by the corresponding generalized inverse. Let
$K:\mathbb R^d\to[0,\infty)$ be a kernel, write $K_h(v)=h^{-d}K(v/h)$
for $h>0$ \textcolor{black}{and} choose a bandwidth $h_N>0$.}
{\color{black}\hypersetup{allcolors=black,citecolor=blue}%
For $\tau\in(0,1)$, the kernel distribution function and
its quantile estimator are
}%
\[
{%
\widehat F_N(u;x)
:=\frac{\sum_{t=L_N+1}^{N}K_{h_N}(X_t-x)\,\1\{R_{N,t}\le u\}}
{\sum_{t=L_N+1}^{N}K_{h_N}(X_t-x)},
\qquad
{\color{black}{\color{black}\widehat q^{\mathrm{ker}}_{N,\tau}}}(x):=\inf\{u:\widehat F_N(u;x)\ge\tau\}.}
\]%
{{\color{black}Through the kernel weights, observations with covariates close to $x$ receive greater emphasis.}
Using the estimated quantiles at levels $\alpha/2$ and
$1-\alpha/2$, we construct a two-sided $(1-\alpha)$ prediction interval
for $Y_{N+1}$ as}
\[
{%
\widehat C_N(X_{N+1})
:=\Bigl[\,
\widehat Y_{N,N+1}+{\color{black}{\color{black}\widehat q^{\mathrm{ker}}_{N,\alpha/2}}}(X_{N+1}),\;
\widehat Y_{N,N+1}+{\color{black}{\color{black}\widehat q^{\mathrm{ker}}_{N,1-\alpha/2}}}(X_{N+1})
\Bigr].}
\]%

\Needspace{6\baselineskip}
{\color{black}A common alternative is to use absolute residuals as conformity
scores.}
{Replacing the signed residual by $A_{N,t}=|R_{N,t}|$ yields an
\textcolor{black}{absolute residual} version of \textcolor{black}{TRACK},
denoted \textcolor{black}{\mbox{TRACK-Abs}}.}
{Let ${\color{black}{\color{black}\widehat q^{\mathrm{ker,abs}}_{N,1-\alpha}}}(x)$ be the estimated
$(1-\alpha)$-quantile obtained from the same kernel
{distribution function} with $R_{N,t}$ replaced by $A_{N,t}$.
Its prediction interval is}
\[
{%
\widehat C_N^{\mathrm{abs}}(X_{N+1})
:=\Bigl[\,
\widehat Y_{N,N+1}-{\color{black}{\color{black}\widehat q^{\mathrm{ker,abs}}_{N,1-\alpha}}}(X_{N+1}),\;
\widehat Y_{N,N+1}+{\color{black}{\color{black}\widehat q^{\mathrm{ker,abs}}_{N,1-\alpha}}}(X_{N+1})
\Bigr].}
\]%
\Needspace{3\baselineskip}

\begingroup
\color{black}\hypersetup{allcolors=black,citecolor=blue}
\begin{remark}
\label{rem:selected-covariates}
Kernel weighting can suffer from the curse of dimensionality. One may
therefore compute the kernel weights using a prespecified selection of
important covariates $S$, while retaining the full $X$ for point prediction.
Under the assumptions stated \textcolor{black}{in Sections~\ref{subsec:finite-guarantees} and~\ref{subsec:conditional-score-quantiles}}, with the covariate conditions imposed
on $S$, the {\color{black}\hypersetup{allcolors=black,citecolor=blue}rolling-window upper bounds and coverage results \textcolor{black}{in Sections~\ref{sec:upper-bounds} and~\ref{sec:conditional-validity}}} remain valid with
$S$ replacing $X$ as the conditioning variable and $\dim(S)$ replacing $d$.
We adopt this approach in the real-data studies in
Section~\ref{sec:real-data} and report the corresponding results.
\end{remark}
\endgroup

\begingroup
\color{black}\hypersetup{allcolors=black,citecolor=blue}
\begin{remark}
\label{rem:general-predictors}
The algorithms also construct prediction sets for general predictors.
Sections~\ref{subsec:general-predictors} and~\ref{subsec:general-conditional-validity} {\color{black}give guarantees when these predictors
are sufficiently close to rolling-window predictors.}
\end{remark}
\par
\endgroup

\subsection{\textcolor{black}{Multi-step}-ahead prediction}\label{sec:multi_step}
{We now turn to \textcolor{black}{multi-step}-ahead \textcolor{black}{prediction} with
horizon $H_N>1$.} In this setting, the objective is to construct forecasts
together with prediction sets for the sequence of future responses over
the next {$H_N$} periods.
Given the information available at time $t$, namely the historical data
$\{Z_i\}_{i=1}^t$ together with the newly observed covariate vector
$X_{t+1}$, {\color{black}we use the collection of point \textcolor{black}{forecasts}
$\{g_{N,h}\}_{h=1}^{H_N}$ to define the $h$-step-ahead forecast as}
\[
{%
    \widehat Y_{N,t+h\mid t}
    =g_{N,h}\!\left(X_{t+1};Z_t,Z_{t-1},\ldots,Z_{t-L_N+1}\right),
    \qquad h\in[H_N],\quad \textcolor{black}{t\in\{L_N,\ldots,N\}},}
\]%
where {$\widehat Y_{N,t+h\mid t}$} denotes the prediction of
$Y_{t+{h}}$ formed at time $t$ using the information available
up to that time. \textcolor{black}{For each $t$}, the collection
{$\{\widehat Y_{N,t+h\mid t}\}_{h=1}^{H_N}$} represents the
sequence of forecasts for the next {$H_N$} periods produced
from a common information set.

\begin{algorithm}[!tbp]
	\caption{\textcolor{black}{\mbox{\textcolor{black}{TRACK-J}}: Joint} \textcolor{black}{Multi-Step}-Ahead Prediction Set}
	\label{alg:multi-step}
	\algorithmspacing
	\textbf{Input:} Data $\{Z_1,\ldots,Z_{N}, X_{{N}+1}\}$, predictors ${\color{black}\{g_{N,h}\}_{h=1}^{H_N}}$, kernel $K$, \mbox{training length ${L_N}$}, \mbox{rolling scale length ${M_N}$}, \mbox{bandwidth $h_{N}$}, significance level $\alpha$.\par
	\textbf{Output:} Prediction set $\widehat{C}_{N}(X_{{N}+1})$.\par
	\begin{enumerate}[label=\arabic*.,leftmargin=1.6em,labelsep=0.55em,topsep=5pt,itemsep=5pt,parsep=0pt,partopsep=0pt]
		\item {\textcolor{black}{For $t=L_N,\ldots,N$ and each $h\in[H_N]$,} compute the $h$-step forecast using $g_{N,h}$:}
		\[
{%
        \widehat Y_{N,t+h\mid t}
        =g_{N,h}\bigl(X_{t+1};Z_t,\ldots,Z_{t-L_N+1}\bigr).
        }%
\]%
		\item Define the errors ${e_{N,t+h\mid t}:=Y_{t+h}-\widehat Y_{N,t+h\mid t}}$ and the score
		\[
{%
        R_{N,t}:=\textcolor{black}{\frac{1}{H_N}}\sum_{j=1}^{H_N}
        \frac{e_{N,t+j-1\mid t-1}^2}
        {\textcolor{black}{\frac{1}{M_N}}\sum_{m=1}^{M_N}e_{N,t-m\mid t-m-j}^2}.
        }%
\]%
		\item \textcolor{black}{Define} the locally weighted empirical \mbox{distribution function}
		\[
{%
		{\widehat F_N}(u;x):=\frac{\sum_{t = {M_N+L_N+H_N}}^{N-H_N+1} K_{h_{N}}(X_t-x)\,\1\{R_{N,t}\le u\}}{\sum_{t = {M_N+L_N+H_N}}^{N-H_N+1} K_{h_{N}}(X_t-x)}.
		}%
\]%
		\item For each $x \in \mathcal X$ and $\tau\in(0,1)$, \textcolor{black}{define}
		\[
		\widehat q^{\mathrm{ker}}_{N,\tau}(x):=\inf\{u:{\widehat F_N}(u;x)\ge\tau\}.
		\]%
		\item {\color{black}\hypersetup{allcolors=black,citecolor=blue}Using $R_{N,N+1}(y)$ from
		{\color{black}\eqref{eq:multistep-candidate-score}}, the prediction set is \textcolor{black}{obtained as}}
		\[
{%
			\widehat C_{{N}}(X_{{N}+1})
			=
			\Big\{y\in\mathbb R^{H_N}:
			R_{N,N+1}(y)\le {\color{black}{\color{black}\widehat q^{\mathrm{ker}}_{N,1-\alpha}}}(X_{{N}+1})\Big\}.
		}%
\]%
	\end{enumerate}
\end{algorithm}

{\textcolor{black}{Multi-step} prediction targets the future vector
$Y_{N+1:N+H_N}=(Y_{N+1},\ldots,Y_{N+H_N})^\top\in\mathbb R^{H_N}$.}
{We compress the vector of forecast errors into a scalar
conformity score for joint calibration.}
{Define the $h$-step-ahead forecast error as
$e_{N,t+h\mid t}:=Y_{t+h}-\widehat Y_{N,t+h\mid t}$,
$h\in[H_N]$, and collect the errors at time $N$ in
$E_N:=(e_{N,N+1\mid N},\ldots,e_{N,N+H_N\mid N})^\top\in\mathbb R^{H_N}$.}

Since smaller absolute components of $E_{N}$ correspond to
more accurate forecasts, a natural summary is $\|E_{N}\|_2^2$.
{However, horizon-dependent error scales can lead this sum to
overweight \textcolor{black}{typically later horizons}. We therefore normalize each horizon using a
rolling \textcolor{black}{average} of squared forecast errors. Fix a rolling scale length
$M_N\in\mathbb N$. For $j\in[H_N]$, define
\begin{equation}\label{eq:normalized-score}
    S_{N,t,j}:=\textcolor{black}{\frac{1}{M_N}}\sum_{m=1}^{M_N}e_{N,t-m\mid t-m-j}^2,
    \qquad
    R_{N,t}:=\textcolor{black}{\frac{1}{H_N}}\sum_{j=1}^{H_N}\frac{e_{N,t+j-1\mid t-1}^2}{S_{N,t,j}}.
\end{equation}%
{\color{black}\hypersetup{allcolors=black,citecolor=blue}\widowpenalty=10000%
Here, \textcolor{black}{$S_{N,t,j}$} is the rolling average of past squared forecast
errors at horizon $j$, which aims to estimate the forecast error variance
at each horizon. \textcolor{black}{Other choices of $S_{N,t,j}$ are also allowed, including unnormalized sums of squared forecast errors and weighted averages. For example, an exponentially weighted average is}
\[
\textcolor{black}{S_{N,t,j}^{(\gamma)}:=
\frac{\sum_{m=1}^{M_N}\gamma^m e_{N,t-m\mid t-m-j}^{2}}
{\sum_{m=1}^{M_N}\gamma^m},\qquad 0<\gamma<1.}
\]
\textcolor{black}{More generally, the framework accommodates a broad class of scale estimators defined as time-invariant measurable functions of a finite window of past observations.}\par
}}%

\Needspace{6\baselineskip}
{\color{black}As in one-step prediction, we consider the conditional distribution
of $R_{N,t}$ given $X_t$. By stationarity, this distribution is time invariant.}
{\color{black}For $\tau\in(0,1)$, its conditional cumulative distribution
function and corresponding quantile are}
\[
{%
    F_N(u\mid x):=\mathbb P\{R_{N,t}\le u\mid X_t=x\},
    \qquad
    {\color{black}q_{N,\tau}(x):=\inf\{u:F_N(u\mid x)\ge\tau\}}.}
\]%
{\color{black}\hypersetup{allcolors=black,citecolor=blue}The score $R_{N,t}$ depends on
$Z_{t-M_N-L_N-H_N+1},\ldots,Z_{t+H_N-1}$. Its full window is observed
within $(Z_1,\ldots,Z_N)$ when
$t-M_N-L_N-H_N+1\ge1$ and $t+H_N-1\le N$. Thus the available scores are
$R_{N,M_N+L_N+H_N},\ldots,R_{N,N-H_N+1}$, indexed by
\begin{equation*}\color{black}
 \mathcal I_N=\{M_N+L_N+H_N,\ldots,N-H_N+1\}.
\end{equation*}%
{\color{black}Calibration on these indices at level $1-\alpha$ yields
$\widehat q^{\mathrm{ker}}_{N,1-\alpha}(X_{N+1})$ for the target score $R_{N,N+1}$.}}
{For a candidate trajectory
$y=(y_1,\ldots,y_{H_N})^\top\in\mathbb R^{H_N}$, define}
{\color{black}%
\begin{equation}\label{eq:multistep-candidate-score}
    R_{N,N+1}(y):=\textcolor{black}{\frac{1}{H_N}}\sum_{j=1}^{H_N}
    \frac{(y_j-\widehat Y_{N,N+j\mid N})^2}{S_{N,N+1,j}}.
\end{equation}}%
{The prediction set contains trajectories whose scores do not
exceed {\color{black}the corresponding} estimated quantile.}
{\hypersetup{linkcolor=black}\textcolor{black}{We denote this joint multi-step extension of TRACK by \mbox{TRACK-J}.
Algorithm~{\color{black}\ref{alg:multi-step}} gives the complete procedure.}}

{The following sections study the calibration error arising
from kernel localization, temporal dependence \textcolor{black}{and} overlap between score
windows. We consider both an independent benchmark target and the actual
future target, \textcolor{black}{and} investigate finite-sample guarantees, conditional
validity \textcolor{black}{and} minimax limits. Throughout, the horizon and rolling-window
{\color{black}lengths may vary with $N$, {\color{black}and the analysis makes explicit
how their growth affects calibration accuracy.}}}
\endgroup%

\par
\color{black}%
\section[Finite-sample prediction guarantees]{Finite-sample prediction guarantees}
\label{sec:upper-bounds}
\begingroup%
\color{black}\hypersetup{allcolors=black,citecolor=blue}%
{%
We now derive finite-sample guarantees for the prediction sets in
Section~{\color{black}\ref{sec:mth}}. {By tracking the raw observations entering each score, we account for
dependence among calibration scores as well as their dependence on the
target score.} We first specify the
available data and admissible estimation rules, then introduce losses
for both one-sided and two-sided score sets.}
{\color{black}\hypersetup{allcolors=black,citecolor=blue}Sections \ref{subsec:quantile-rules-losses} and \ref{subsec:finite-guarantees} concern rolling-window predictors, while
Section~\ref{subsec:general-predictors} extends the guarantees to more general predictors.}

\begin{definition}[\textcolor{black}{Score descriptor and induced \textcolor{black}{row of scores}}]
\label{def:score-row}
For each {raw sample} size $N$, the \emph{score descriptor} is
$\vartheta_N=(a_N,b_N,\rho_N)$, where $a_N,b_N\in\mathbb N_0$,
$D_N=a_N+b_N<N$ \textcolor{black}{and}
$\rho_N:\mathcal Z^{D_N+1}\to\mathbb R$ {is a known time-invariant
measurable map.}
The integers $a_N$ and $b_N$ specify the backward and forward extents
of the score window relative to $t$.
{We call $D_N$ the {\textbf{\textit{\textcolor{black}{score window} span}}}.} The induced score is
{\begin{equation*}
 R_{N,t}=\rho_N(Z_{t-a_N},\ldots,Z_{t+b_N}),\qquad t\ge a_N+1.
\end{equation*}}%
{{\color{black}\hypersetup{allcolors=black,citecolor=blue}With $N$ \textcolor{black}{given}, we abbreviate $R_{N,t}$ as $R_t$. In particular,
the target score is $R_{N+1}=R_{N,N+1}$.}}
\end{definition}

{%
\textcolor{black}{For \textcolor{black}{TRACK}} (Algorithm~{\color{black}\ref{alg:quantile_estimation}}), the descriptor is determined
by the training length $L_N$ and forecasting rule $g_{N,1}$:
$(a_N,b_N)=(L_N,0)$ and
\[
 \rho_N(z_1,\ldots,z_{L_N+1})
 =y_{L_N+1}-g_{N,1}(x_{L_N+1};z_{L_N},\ldots,z_1).
\]}%
\textcolor{black}{\mbox{TRACK-Abs}} replaces this map by its absolute value.
{%
\textcolor{black}{For \mbox{\textcolor{black}{TRACK-J}}} (Algorithm~{\color{black}\ref{alg:multi-step}}), the descriptor also depends on the
forecast horizon $H_N$ and rolling scale length $M_N$, with
\begin{equation*}
 (a_N,b_N)=(M_N+L_N+H_N-1,\ H_N-1).
\end{equation*}%
Thus $D_N=M_N+L_N+2H_N-2$, and the score map is
\[
\begin{aligned}
 &\rho_N(z_1,\ldots,z_{D_N+1})\\
 {}\mathrel{\textcolor{black}{=}}{}&\textcolor{black}{\frac{M_N}{H_N}}\sum_{j=1}^{H_N}
 \frac{\{y_{a_N+j}-g_{N,j}(x_{a_N+1};z_{a_N},\ldots,z_{a_N-L_N+1})\}^2}
 {\sum_{m=1}^{M_N}
 \{y_{a_N+1-m}-g_{N,j}(x_{a_N+2-m-j};
 z_{a_N+1-m-j},\ldots,z_{a_N+2-m-j-L_N})\}^2}.
\end{aligned}
\]}%
\par\noindent {Thus $\vartheta_N$ specifies the scoring procedure at sample size $N$
and is not part of the unknown data-generating law. {\color{black}\hypersetup{allcolors=black,citecolor=blue}Any algorithmic randomness is independent of the raw
process, with the precise conventions given in
the Appendix.}}
{For each fixed $N$, the process
$\{(X_t,R_{N,t})\}_{t\ge a_N+1}$ is stationary in $t$.}
{As $N$ varies, the same underlying process can induce different score
distributions because $\rho_N$ and its window lengths may change.}

By Definition~{\color{black}\ref{def:score-row}}, the full window of $R_{N,t}$ is observed
when $t-a_N\ge1$ and $t+b_N\le N$. We therefore use the calibration
indices $\mathcal I_N=\{a_N+1,\ldots,N-b_N\}$, with
$n_N:=|\mathcal I_N|=N-D_N$.
{Write the observed raw block and its generated sigma-field as
\begin{equation}\label{eq:observed-raw-block}
 O_N:=(Z_1,\ldots,Z_N){\in\mathcal Z^N=\mathbb R^{(d+1)N}},
 \qquad \mathcal G_N:=\sigma(O_N).
\end{equation}%
Every calibration score is $\mathcal G_N$-measurable because its full
window lies in $O_N$. The target score uses
$Z_{N+1-a_N},\ldots,Z_{N+1+b_N}$ and can be evaluated after the future
responses are observed: through time $N+1$ for \textcolor{black}{\textcolor{black}{TRACK} and \mbox{\textcolor{black}{TRACK-Abs}}}
and $N+H_N$ for \textcolor{black}{\mbox{TRACK-J}}. The prediction sets
themselves are constructed from $O_N$ and $X_{N+1}$.}

\subsection[Quantile estimators and evaluation criteria]{Quantile estimators and evaluation criteria}
\label{subsec:quantile-rules-losses}
\textcolor{black}{Fix a data-generating law $P$ and a row of scores $\{R_{N,t}\}_{t\ge a_N+1}$.} For $\tau\in(0,1)$, write
\[
 F_N(u\mid x)=P\{R_{N,t}\le u\mid X_t=x\},\qquad
 q_{N,\tau}(x)=\inf\{u:F_N(u\mid x)\ge\tau\}.
\]%
{%
For a realization {$\textcolor{black}{\mathbf z}=(z_1,\ldots,z_N)\in\mathcal Z^N$}, with
$z_t=(x_t,y_t)$, the calibration data form the ordered array
\begin{equation}\label{eq:score-array-map}
 \mathbf w_{N,\vartheta_N}(\textcolor{black}{\mathbf z})
 :=\bigl(x_t,\rho_N(z_{t-a_N},\ldots,z_{t+b_N})\bigr)_{t\in\mathcal I_N}
 \in\mathcal Z^{n_N}.
\end{equation}%
Let $V\sim\operatorname{Unif}(0,1)$ be a random seed independent of the
raw process. {Let $\mathcal A_N$ be a \textcolor{black}{conditional quantile} estimation
rule, with estimates $\widehat q=(\widehat q_{N,\tau})_{\tau\in(0,1)}$
given by}
\begin{equation}\label{eq:generic-score-quantile-rule}
 \begin{aligned}
 \mathcal A_N&:\mathcal Z^{n_N}\times\mathbb R^d\times(0,1)\times[0,1]
 \longrightarrow\mathbb R,\\
 \widehat q_{N,\tau}(x)
 &=\mathcal A_N\bigl(\mathbf w_{N,\vartheta_N}(O_N),x,\tau;V\bigr).
 \end{aligned}
\end{equation}%
{Since quantiles are nondecreasing in their probability level, we
require the same property of $\mathcal A_N$ in $\tau$ and assume
joint measurability in all its inputs. For each fixed $x$ and $\tau$,
$\widehat q_{N,\tau}(x)$ is $\mathcal G_N\vee\sigma(V)$-measurable.
A deterministic rule ignores $V$ and gives a $\mathcal G_N$-measurable
estimator. {For randomized rules, probabilities and expectations
also average over $V$.} We collect these estimators in}
{\begin{equation}\label{eq:admissible-quantile-class}
 \mathcal D_N(\vartheta_N)
 :=\left\{\widehat q:
 \widehat q_{N,\tau}(x)
 =\mathcal A_N(\mathbf w_{N,\vartheta_N}(O_N),x,\tau;V)\right\}.
\end{equation}}%

{\textcolor{black}{\textcolor{black}{TRACK} and its variants}
use kernel weights to estimate the conditional distribution of
their respective scores. On the available calibration indices,
the distribution function and its quantile estimator are
\begin{equation}\label{eq:kernel-quantile-estimator}
 \begin{aligned}
 \widehat F_N(u;x)
 &:=\frac{\sum_{t\in\mathcal I_N}K_{h_N}(X_t-x)\,
              \1\{R_{N,t}\le u\}}
            {\sum_{t\in\mathcal I_N}K_{h_N}(X_t-x)},\\
 \widehat q^{\mathrm{ker}}_{N,\tau}(x)
 &:=\inf\{u:\widehat F_N(u;x)\ge\tau\}.
 \end{aligned}
\end{equation}%
}%

For nominal coverage $\tau\in(0,1)$, let $C^*_{N,\tau}(x)$ denote
either of the following oracle intervals:
{\begin{equation}\label{eq:oracle-intervals}
 C^*_{N,\tau}(x)
 =(-\infty,q_{N,\tau}(x)]
 \quad\text{or}\quad
 [q_{N,(1-\tau)/2}(x),q_{N,(1+\tau)/2}(x)].
\end{equation}}%
{\emergencystretch=1em At $\tau=1-\alpha$, \textcolor{black}{TRACK} uses
the two-sided \textcolor{black}{score interval}, translated by its point forecast. \textcolor{black}{\mbox{\textcolor{black}{TRACK-Abs}} and \mbox{\textcolor{black}{TRACK-J}}}
use the one-sided \textcolor{black}{score set} at the same nominal coverage.
For an estimator $\widehat q\in\mathcal D_N(\vartheta_N)$, the
corresponding estimated interval is
\[
 \widehat C_{N,\tau}(x)
 =(-\infty,\widehat q_{N,\tau}(x)]
 \quad\text{or}\quad
 [\widehat q_{N,(1-\tau)/2}(x),\widehat q_{N,(1+\tau)/2}(x)].
\]%
Unless a form is specified, the results below apply to both choices
in {\color{black}\eqref{eq:oracle-intervals}}, with the estimated interval of the
same form.}}

To evaluate these sets, we distinguish learning the conditional score
law from predicting the next score in the same series. {In the independent experiment, the calibration observations retain
their original temporal dependence, while the target is independent of
that sample. Using the actual future target also accounts for its
dependence on the calibration history.}

{%
For $\widehat q\in\mathcal D_N(\vartheta_N)$, take a pair
$(X_{N+1}^\circ,R_{N+1}^\circ)$ with the same law as
$(X_t,R_{N,t})$, independent of the raw process and $V$.
The independent-target loss is
}%
{\begin{equation}\label{eq:probability-scale-loss}
 \mathcal L^{\mathrm{ind}}_{P,N,\tau}(\widehat q)
 =P\!\left\{R_{N+1}^\circ\in
 C^*_{N,\tau}(X_{N+1}^\circ)\mathbin\triangle
 \widehat C_{N,\tau}(X_{N+1}^\circ)\right\}.
\end{equation}}%
\Needspace{4\baselineskip}
For the dependent experiment, keep $(X_{N+1},R_{N+1})$ under its actual
joint law with the calibration history and define
{\begin{equation}\label{eq:dependent-set-loss}
 \mathcal L^{\mathrm{dep}}_{P,N,\tau}(\widehat q)
 =P\!\left\{R_{N+1}\in
 C^*_{N,\tau}(X_{N+1})\mathbin\triangle
 \widehat C_{N,\tau}(X_{N+1})\right\}.
\end{equation}}%
Both losses measure the probability that the estimated and oracle sets
disagree on target membership.
{\color{black}Similar criteria based on symmetric differences between
estimated and oracle sets have also been used in conformal prediction
{\color{blue}\citep{lei2018distribution,matabuena2025model}}.}
{{To assess this error at a
fixed covariate value, define the corresponding conditional losses:}
\begin{align}
 \mathcal L^{\mathrm{ind}}_{P,N,\tau}(\widehat q\mid x)
 &:=P\!\left\{R_{N+1}^{\circ}\in
 C^*_{N,\tau}(x)\mathbin\triangle\widehat C_{N,\tau}(x)
 \,\middle|\,X_{N+1}^{\circ}=x\right\},
 \notag\\
 \mathcal L^{\mathrm{dep}}_{P,N,\tau}(\widehat q\mid x)
 &:=P\!\left\{R_{N+1}\in
 C^*_{N,\tau}(x)\mathbin\triangle\widehat C_{N,\tau}(x)
 \,\middle|\,X_{N+1}=x\right\}.
 \label{eq:conditional-set-loss}
\end{align}%
{{The corresponding unconditional losses are obtained by integrating
over $P_X$.} In the independent experiment,
conditioning on the query leaves the calibration law unchanged.
In the dependent experiment, it retains the relationship between
the observed history and the actual future target.}

To connect these losses with the coverage objective in {\color{black}\eqref{eq:between}}, define}

{\begin{equation}\label{eq:coverage-functional}
 \mathcal E_{N,\tau}(\widehat q)
 =\sup_{\|f\|_\infty\le1}
 \left|\mathbb E_P\!\left[
 f(X_{N+1})
 \{\1\{R_{N+1}\in\widehat C_{N,\tau}(X_{N+1})\}-\tau\}
 \right]\right|.
\end{equation}}%
{%
The next lemma bounds this \textcolor{black}{coverage discrepancy} by the probability of
disagreement with the oracle set.
}%

\Needspace{4\baselineskip}
{%
\begin{lemma}[Coverage from set disagreement]
\label{lem:coverage-by-disagreement}
Fix $\tau\in(0,1)$ and $\widehat q\in\mathcal D_N(\vartheta_N)$.
{Choose either oracle interval in {\color{black}\eqref{eq:oracle-intervals}}
and its estimated counterpart.} If $F_N(\cdot\mid x)$ is continuous
for $P_X$-almost every $x$, then
\begin{equation}\label{eq:coverage-dominated-by-set-loss}
 \begin{aligned}
 \bigl|P\{R_{N+1}\in\widehat C_{N,\tau}(x)\mid X_{N+1}=x\}-\tau\bigr|
 &\le\mathcal L^{\mathrm{dep}}_{P,N,\tau}(\widehat q\mid x),\\
 \bigl|P\{R_{N+1}\in\widehat C_{N,\tau}(X_{N+1})\}-\tau\bigr|
 &\le\mathcal L^{\mathrm{dep}}_{P,N,\tau}(\widehat q).
 \end{aligned}
\end{equation}%
The first inequality holds for $P_X$-almost every $x$.
{Both inequalities also hold in the independent experiment, with
$(X_{N+1},R_{N+1})$ replaced by
$(X_{N+1}^{\circ},R_{N+1}^{\circ})$ and
$\mathcal L^{\mathrm{dep}}$ by $\mathcal L^{\mathrm{ind}}$.}
{In addition,}
\[
 {\mathcal E_{N,\tau}(\widehat q)
 \le\mathcal L^{\mathrm{dep}}_{P,N,\tau}(\widehat q).}
\]%
\end{lemma}

{%
The conditional and unconditional losses control conditional and
marginal coverage respectively. Translating a \textcolor{black}{signed residual}
interval or taking the preimage of an \textcolor{black}{absolute residual} or \textcolor{black}{multi-step} score
set preserves target membership. The analysis below therefore covers
\textcolor{black}{\textcolor{black}{TRACK} and its variants}.
}}%

\subsection[Assumptions and finite-sample guarantees]{Assumptions and finite-sample guarantees}
\label{subsec:finite-guarantees}
Having specified the prediction sets and their evaluation criteria, we now
establish theoretical guarantees for \textcolor{black}{these procedures}. Guided by
the coverage objective in {\color{black}\eqref{eq:between}}, we first bound the discrepancy
between actual and nominal coverage through \textcolor{black}{score set} disagreement.
Our analysis is conducted under the following regularity conditions.

\begin{assumption}
\label{ass:distribution}
{%
The law of the raw process $\{Z_t=(X_t,Y_t)\}_{t\in\mathbb N}$
satisfies the following conditions.
}%
\begin{enumerate}[label=(\roman*)]
 \item The process $Z_t=(X_t,Y_t)$ is strictly stationary and absolutely
 regular \textcolor{black}{with $\beta$-mixing coefficients satisfying}
 \[
  \beta_Z(k)\le C_\beta k^{-p},\qquad k\ge1,
 \]%
 for constants $C_\beta<\infty$ and $p>2$.
 \item The support $\mathcal X=\operatorname{supp}(P_X)\subset\mathbb R^d$
 is compact and, for some $\kappa_X,r_X>0$,
 \[
  \operatorname{Leb}\{\mathcal X\cap B(x,r)\}
  \ge\kappa_X\operatorname{Leb}\{B(x,r)\},\quad
  {\color{black}\hypersetup{allcolors=black,citecolor=blue}\forall x\in\mathcal X,\quad 0<r\le r_X.}
 \]%
 \item {
The marginal law of $X_t$ and the joint laws of
 $(X_1,X_{1+k})$, $k\ge\nu$, have Lebesgue densities $f_X$ and
 $f_{X_1,X_{1+k}}$ satisfying
 \[
  {f_X(x)\ge c_X>0},\qquad
  \sup_{k\ge\nu} f_{X_1,X_{1+k}}(x,y)\le C_X,
 \]%
 for every $x,y\in\mathcal X$, for some fixed $\nu\in\mathbb N$
 and constants $c_X>0$ and $C_X<\infty$.
}%
\end{enumerate}
\end{assumption}

{\color{black}\hypersetup{allcolors=black,citecolor=blue}%
Definitions of strict stationarity and absolute regularity are given in
the Appendix. Condition~(ii) is standard in
nonparametric minimax theory {\color{blue}\citep{audibert2007fast}}, while related mixing
and density conditions appear in time series kernel estimation
{\color{blue}\citep{hansen2008uniform}}. The restriction $k\ge\nu$ in (iii) accommodates
covariates formed from overlapping lagged observations. Conditions~(ii)--(iii)
also imply that $f_X$ is bounded above.
{\color{black}\hypersetup{allcolors=black,citecolor=blue}%
The following example gives a familiar model satisfying these conditions.}}

\begingroup
\color{black}\hypersetup{allcolors=black,citecolor=blue}
\begin{example}
\label{ex:ar-distribution}
Consider $Z_t=(X_t,Y_t)$ with iid covariates
$X_t\sim\operatorname{Unif}[-1,1]$ and
\[
 Y_t=\theta_0+\theta_1X_t+U_t,\qquad U_t=\phi U_{t-1}+\xi_t,
\]
where $|\phi|<1$, $\xi_t\overset{\mathrm{iid}}{\sim}N(0,\sigma^2)$ with
$\sigma>0$, and $U_t$ is a stationary Gaussian AR(1) process. Assume the
covariate and innovation sequences are independent. Then
\[
 \beta_Z(k)\le\frac{|\phi|^k}{2\sqrt{1-\phi^2}},\qquad k\ge1.
\]
Hence $\beta_Z(k)=O(k^{-p})$ for every $p>0$, and
Assumption~\ref{ass:distribution} holds.
\end{example}
\endgroup

\begin{assumption}
\label{ass:conditional-score-cdf}
For each $N$ and $x\in\mathcal X$, $u\mapsto F_N(u\mid x)$ is continuous.
For some $\zeta\in(0,1]$ and $L_F<\infty$ independent of $N$,
\begin{equation*}
 \sup_{u\in\mathbb R}|F_N(u\mid x)-F_N(u\mid x')|
 \le L_F\|x-x'\|^\zeta,\quad
 {\color{black}\hypersetup{allcolors=black,citecolor=blue}\forall x,x'\in\mathcal X,\quad \forall N\ge1.}
\end{equation*}%
\end{assumption}

{\color{black}\hypersetup{allcolors=black,citecolor=blue}%
Assumption~\ref{ass:conditional-score-cdf} controls the local smoothing bias
through smoothness of the conditional score distribution across covariates,
a condition also used in localized conformal prediction {\color{blue}\citep{guan2023localized}}.
We require the smoothness bound to hold uniformly in $N$, since the score
distribution may vary with $N$.
{\color{black}\hypersetup{allcolors=black,citecolor=blue}%
The following example verifies this condition for ordinary least squares
(OLS) fitted on the most recent $L_N$ observations.}}

\begingroup
\color{black}\hypersetup{allcolors=black,citecolor=blue}
\begin{example}
\label{ex:ar-ols}
In the model of Example~\ref{ex:ar-distribution}, take deterministic
$3\le L_N<N$ and define
\[
 (\widehat\theta_{0,N,t},\widehat\theta_{1,N,t})
 =\mathop{\arg\min}_{(a,b)\in\mathbb R^2}
       \sum_{i=1}^{L_N}(Y_{t-i}-a-bX_{t-i})^2.
\]
Set $\widehat Y_{N,t}=\widehat\theta_{0,N,t}+\widehat\theta_{1,N,t}X_t$.
The signed residuals $R_{N,t}=Y_t-\widehat Y_{N,t}$ satisfy
Assumption~\ref{ass:conditional-score-cdf} with
\[
 \zeta=1,\qquad
 L_F=\frac{3}{\pi\sqrt2(1-|\phi|)}.
\]
Their conditional densities are bounded by $1/(\sigma\sqrt{2\pi})$,
uniformly in $N$ and $L_N$. For absolute residuals, the H\"older constant
and density bound both double.
\end{example}
\endgroup

\begin{assumption}
\label{ass:kernel}
For the exponent $\zeta$ in Assumption~{\color{black}\ref{ass:conditional-score-cdf}},
the kernel $K:\mathbb R^d\to[0,\infty)$ is bounded and satisfies
\[
 \int K(v)\,dv=1,\qquad
 \int\|v\|^\zeta K(v)\,dv<\infty.
\]%
Moreover, $K(0)>0$ and $K$ is continuous on a neighborhood of the origin.
\end{assumption}

{\color{black}\hypersetup{allcolors=black,citecolor=blue}%
Assumption~\ref{ass:kernel} imposes standard kernel conditions.
Under Assumptions~\ref{ass:distribution}--\ref{ass:kernel}, the following
theorem gives finite-sample bounds on prediction set disagreement for
the kernel conditional quantile estimator in
\eqref{eq:kernel-quantile-estimator}.}

\Needspace{8\baselineskip}
\begin{theorem}[\textcolor{black}{Bounds for prediction set disagreement}]
\label{thm:overlap-upper}
Fix $\tau_0\in(0,1/2)$ \textcolor{black}{and} suppose
Assumptions~{\color{black}\ref{ass:distribution}}--{\color{black}\ref{ass:kernel}} hold for the \textcolor{black}{row of scores}
induced by $\vartheta_N=(a_N,b_N,\rho_N)$. For the kernel \textcolor{black}{conditional quantile}
estimator $\widehat q^{\mathrm{ker}}$ in {\color{black}\eqref{eq:kernel-quantile-estimator}}
with bandwidth $h_N$, there exist constants $h_0\in(0,1]$ and $C>0$,
independent of $N,h_N,\vartheta_N,\tau$, such that, for
$0<h_N\le h_0$ and $\tau\in[\tau_0,1-\tau_0]$,
{\small%
{\color{black}\hypersetup{allcolors=black,citecolor=blue}
\begin{equation}\label{eq:finite-sample-upper-main}
 \mathcal L^{\mathrm{ind}}_{P,N,\tau}(\widehat q^{\mathrm{ker}})
 \vee\mathcal L^{\mathrm{dep}}_{P,N,\tau}(\widehat q^{\mathrm{ker}})
 \le C\left\{h_N^\zeta+
 \sqrt{\frac{h_N^{-d}+D_N}{n_N}}\right\}.
\end{equation}
}%
}%
If $h_N\asymp n_N^{-1/(2\zeta+d)}$, then
{\small%
{\color{black}\hypersetup{allcolors=black,citecolor=blue}
\begin{equation}\label{eq:optimized-upper-main}
 \mathcal L^{\mathrm{ind}}_{P,N,\tau}(\widehat q^{\mathrm{ker}})
 \vee\mathcal L^{\mathrm{dep}}_{P,N,\tau}(\widehat q^{\mathrm{ker}})
 \le C\left\{n_N^{-\zeta/(2\zeta+d)}+
 \sqrt{\frac{D_N}{n_N}}\right\}.
\end{equation}
}%
}%
{Both bounds hold for either the one-sided or the two-sided interval
in {\color{black}\eqref{eq:oracle-intervals}}.}
\end{theorem}

\Needspace{4\baselineskip}
\begin{remark}
\label{rem:algorithm-rates}
Substituting the descriptors after Definition~{\color{black}\ref{def:score-row}} into
{\color{black}\eqref{eq:optimized-upper-main}} gives bounds of orders
{%
\[
 \begin{aligned}
 \text{\textcolor{black}{TRACK}:}\quad
 &n_N^{-\zeta/(2\zeta+d)}+\sqrt{\frac{L_N}{n_N}},
 \qquad n_N=N-L_N,\\[5pt]
 \text{\textcolor{black}{\mbox{TRACK-J}}:}\quad
 &n_N^{-\zeta/(2\zeta+d)}+\sqrt{\frac{M_N+L_N+2H_N}{n_N}},
 \qquad n_N=N-M_N-L_N-2H_N+2,
 \end{aligned}
\]%
}%
for all losses covered by Theorem~{\color{black}\ref{thm:overlap-upper}}.
The first rate also applies to \textcolor{black}{\mbox{TRACK-Abs}}.
{%
For \textcolor{black}{these procedures}, the implicit constants are uniform in $N,L_N,M_N,H_N$.}
\end{remark}

For the actual future target, the following corollary transfers the
dependent loss bound to coverage within measurable covariate groups
for the response sets reported by \textcolor{black}{these procedures}.

\Needspace{4\baselineskip}
\begin{corollary}[\textcolor{black}{Coverage within covariate groups}]
\label{cor:group_coverage}
\emergencystretch=3em
Fix $\alpha\in(0,1)$ and consider \textcolor{black}{\textcolor{black}{TRACK} or either of its variants}. Suppose
Assumptions~{\color{black}\ref{ass:distribution}}--{\color{black}\ref{ass:kernel}} hold for the
corresponding descriptor $\vartheta_N$, and use
$0<h_N\le h_0$ with $h_N\asymp n_N^{-1/(2\zeta+d)}$.
Then, for every measurable $A\subseteq\mathcal X$ with $P_X(A)>0$,
\[
 \left|P\{Y_{N+1:N+H_N}\in\widehat C_N(X_{N+1})\mid X_{N+1}\in A\}
 -(1-\alpha)\right|
 \le\frac{C}{P_X(A)}
 \left\{n_N^{-\zeta/(2\zeta+d)}+\sqrt{\frac{D_N}{n_N}}\right\}.
\]%
Here $H_N=1$ for one-step prediction. {The constant $C$ is
independent of $N,\vartheta_N,A$.}
\end{corollary}

{%
Taking $A=\mathcal X$ recovers marginal coverage. For smaller groups,
the factor $1/P_X(A)$ quantifies the cost of localization, so shrinking
$A$ around a covariate value does not directly yield a pointwise guarantee.
{\color{black}\hypersetup{allcolors=black,citecolor=blue}Section~\ref{sec:conditional-validity} studies conditional coverage at a fixed covariate value}
\textcolor{black}{through the behavior of the estimated score quantiles,
with explicit bounds on the \textcolor{black}{coverage discrepancy}.}
}%

\begingroup
\color{black}\hypersetup{allcolors=black,citecolor=blue}
\subsection{Extension to general predictors}
\label{subsec:general-predictors}
\textcolor{black}{Under stationarity, fitting a model to all past observations may yield better forecasts.
These forecasts may actually be close to those obtained from the most recent $L_N$ observations when $L_N$ is suitably large.} To extend the
preceding guarantees to this setting, we consider the following class
of predictors for TRACK and TRACK-Abs. The extension to TRACK-J is
given in the Appendix.

\begin{definition}[\textcolor{black}{Rolling-window approximable predictors}]
\label{def:predictor-approximation}
Measurable predictors $\widetilde g_{N,1,t}$ are \emph{rolling-window
approximable} with error $\varepsilon_N\ge0$ and failure probability
$\pi_N\in[0,1]$ if, for some {\color{black}$L_N\in\{0,\ldots,N\}$},
there are measurable rules $g_{N,1}$ as in \eqref{eq:point_forecast} satisfying
\begin{equation}\label{eq:predictor-approximation}
 P\left\{
 \max_{L_N\le t\le N}
 \left\|\widetilde g_{N,1,t}
 -g_{N,1}(\,\cdot\,;Z_t,\ldots,Z_{t-L_N+1})\right\|_\infty
 >\varepsilon_N\right\}\le\pi_N.
\end{equation}
\end{definition}

\Needspace{4\baselineskip}
\begin{example}\label{ex:expanding-predictor}
In the model of Example~\ref{ex:ar-distribution}, let
$\widetilde g_{N,1,t}$ be OLS fitted to observations $1,\ldots,t$.
If $\log N\ll L_N\ll N$, the rolling OLS predictor in
Example~\ref{ex:ar-ols} gives the approximation in
Definition~\ref{def:predictor-approximation} for all sufficiently
large $N$, with
\[
 \varepsilon_N=C\sqrt{\frac{\log N}{L_N}},
 \qquad \pi_N=N^{-2},
\]
where $C$ depends only on $\sigma$ and $|\phi|$.
\end{example}

We denote the empirical quantile, score interval and oracle interval
associated with $\widetilde g_{N,1,t}$ by $\widetilde q^{\mathrm{ker}}_{N,\tau}$,
$\widetilde C_{N,\tau}$ and $\widetilde C^*_{N,\tau}$ respectively.
Write $\widetilde R_{N+1}:=\widetilde R_{N,N+1}$ for the target score
and $\widetilde C_N$ for the prediction set for $Y_{N+1}$.
We retain the original notation for the corresponding rolling-window
approximation $g_{N,1}$ and impose the following density condition on its scores.

\begin{assumption}\label{ass:comparison-density}
The conditional density of the scores generated by $g_{N,1}$ satisfies
\[
 \sup_N\sup_{x\in\mathcal X}\sup_{u\in\mathbb R}f_N(u\mid x)
 \le B<\infty.
\]
\end{assumption}

The approximation in Definition~\ref{def:predictor-approximation} and the
density bound in Assumption~\ref{ass:comparison-density} control the
additional calibration error. The following
proposition extends the bounds in Theorem~\ref{thm:overlap-upper} to
these predictors.

\begin{proposition}[Bounds for general predictors]
\label{cor:finite-approximation}
Fix $\tau_0\in(0,1/2)$. For TRACK and TRACK-Abs, consider rolling-window approximable
predictors $\widetilde g_{N,1,t}$ with an approximation $g_{N,1}$
satisfying Assumptions~\ref{ass:distribution}--\ref{ass:comparison-density}.
Take $h_N\asymp n_N^{-1/(2\zeta+d)}$.
Let $(X_{N+1}^{\circ},\widetilde R_{N+1}^{\circ})$ be an independent copy of
$(X_{N+1},\widetilde R_{N+1})$ as in Section~\ref{subsec:quantile-rules-losses}.
There exists $C>0$, independent of $N,h_N,\vartheta_N,\tau$,
such that, for $\tau\in[\tau_0,1-\tau_0]$,
\[
 P\!\left\{\widetilde R_{N+1}^{\circ}\in
 \widetilde C^*_{N,\tau}(X_{N+1}^{\circ})\mathbin\triangle
 \widetilde C_{N,\tau}(X_{N+1}^{\circ})\right\}
 \le C\left\{n_N^{-\zeta/(2\zeta+d)}+
 \sqrt{\frac{D_N}{n_N}}+\varepsilon_N+\pi_N\right\},
\]
and
\begin{equation}\label{eq:reference-transfer-optimized}
 P\!\left\{\widetilde R_{N+1}\in
 \widetilde C^*_{N,\tau}(X_{N+1})\mathbin\triangle
 \widetilde C_{N,\tau}(X_{N+1})\right\}
 \le C\left\{n_N^{-\zeta/(2\zeta+d)}+
 \sqrt{\frac{D_N}{n_N}}+\varepsilon_N+\pi_N\right\}.
\end{equation}
Both bounds hold for either interval form in \eqref{eq:oracle-intervals}.
\end{proposition}
Allowing more flexible predictors adds $\varepsilon_N+\pi_N$ to the
bounds in Theorem~\ref{thm:overlap-upper}, where the two terms account
for approximation error and failure probability respectively.
\par
\endgroup

\endgroup%

\section[Quantile estimation and conditional coverage]{\textcolor{black}{Quantile estimation and conditional coverage}}
\label{sec:conditional-validity}
\begingroup%
\color{black}%
\hypersetup{allcolors=black,citecolor=blue}%
{\color{black}\hypersetup{allcolors=black,citecolor=blue}
The finite-sample bounds in Section~\ref{sec:upper-bounds}
lead to two questions: whether the estimated score quantiles converge
to their population counterparts and whether the resulting prediction
sets attain nominal conditional coverage. We first study quantile
estimation, then conditional coverage at the future covariate value.%
}
{\color{black}\hypersetup{allcolors=black,citecolor=blue}Sections \ref{subsec:conditional-score-quantiles} and \ref{subsec:conditional-coverage-bounds} concern rolling-window predictors, while
Section~\ref{subsec:general-conditional-validity} {\color{black}extends these results} to more general predictors.}

\Needspace{6\baselineskip}
\subsection[Conditional score quantile estimation]{\textcolor{black}{Conditional score quantile estimation}}\label{subsec:conditional-score-quantiles}
{\color{black}\hypersetup{allcolors=black,citecolor=blue}Write $f_N(\cdot\mid x)$ for the local conditional score density
in the assumption below. To identify the target quantile, we require
this density to be bounded away from zero in a neighborhood of the quantile.}

{\color{black}\hypersetup{allcolors=black,citecolor=blue}%
\begin{assumption}
\label{ass:local-density}
For any fixed $\tau\in(0,1)$ and every $x\in\mathcal X$, there exist
$c_{\tau,x},\eta_{\tau,x}>0$ such that, for all sufficiently large $N$,
{\color{black}\hypersetup{allcolors=black,citecolor=blue}the conditional score law given $X_t=x$ has a local density
$f_N(\cdot\mid x)$ satisfying $f_N(u\mid x)\ge c_{\tau,x}$ for every $u$
with $|u-q_{N,\tau}(x)|\le\eta_{\tau,x}$.}
\end{assumption}
}%

{\color{black}\hypersetup{allcolors=black,citecolor=blue}
Under the local identification condition,
Theorem~\ref{thm:quantile-consistency} bounds quantile estimation
errors at deterministic covariate values and at the potentially
dependent future covariate.
}

{\color{black}\hypersetup{allcolors=black,citecolor=blue}%
\Needspace{4\baselineskip}
{\color{black}\hypersetup{allcolors=black,citecolor=blue}
\begin{theorem}[Bounds for conditional score quantiles]
\label{thm:quantile-consistency}
Fix $\tau\in(0,1)$. Suppose Assumptions~\ref{ass:distribution}--\ref{ass:kernel}
and \ref{ass:local-density} hold.
There exists a constant $h_0\in(0,1]$ such that, for every fixed
$\varepsilon>0$, there is $C_{\tau,\varepsilon}>0$, independent of
$x,N,h_N$, for which the following bounds hold for $0<h_N\le h_0$:
\begin{equation}\label{eq:quantile-value-consistency}
\begin{gathered}
 \sup_{x\in\mathcal X}P\!\left\{
 |\widehat q^{\mathrm{ker}}_{N,\tau}(x)-q_{N,\tau}(x)|>\varepsilon\right\}
 \le C_{\tau,\varepsilon}\left\{h_N^\zeta+\sqrt{\frac{h_N^{-d}+D_N}{n_N}}\right\},\\
 P\!\left\{|\widehat q^{\mathrm{ker}}_{N,\tau}(X_{N+1})-q_{N,\tau}(X_{N+1})|>\varepsilon\right\}
 \le C_{\tau,\varepsilon}\left\{h_N^\zeta+\sqrt{\frac{h_N^{-d}+D_N}{n_N}}\right\}.
\end{gathered}
\end{equation}
If $h_N\asymp n_N^{-1/(2\zeta+d)}$, the expression in braces
can be replaced by $n_N^{-\zeta/(2\zeta+d)}+\sqrt{D_N/n_N}$.
\end{theorem}
}
}%

{\color{black}\hypersetup{allcolors=black,citecolor=blue}
Convergence at a deterministic $x$, even with an error probability
bound uniform in $x$, does not by itself imply convergence at $X_{N+1}$,
which may depend on the observed history.
Assumption~\ref{ass:local-density} is used to bound quantile estimation errors.
The bounds in Section~\ref{sec:upper-bounds}, the conditional coverage
results below and the minimax rates in Section~\ref{sec:minimax}
do not require this assumption.
}

\Needspace{6\baselineskip}
\subsection{\textcolor{black}{Conditional coverage bounds}}\label{subsec:conditional-coverage-bounds}
{\color{black}\hypersetup{allcolors=black,citecolor=blue}{\color{black}\hypersetup{allcolors=black,citecolor=blue}Coverage over covariate groups of positive probability, as established in
Corollary~{\color{black}\ref{cor:group_coverage}}, does not determine coverage at an
individual covariate value. Pointwise conditional coverage instead fixes
$X_{N+1}=x$ and averages over the remaining randomness.
Theorem~{\color{black}\ref{thm:conditional-coverage}} establishes this guarantee for the
actual future target and gives a uniform version for the independent
benchmark.}}

{\color{black}\hypersetup{allcolors=black,citecolor=blue}%
{\color{black}\hypersetup{allcolors=black,citecolor=blue}
\begin{theorem}[Bounds for conditional coverage]
\label{thm:conditional-coverage}
Suppose Assumptions~\ref{ass:distribution}--\ref{ass:kernel} hold.
There exists a constant $h_0\in(0,1]$ such that the following statements
hold for either interval form in \eqref{eq:oracle-intervals}.
\begin{enumerate}[label=(\roman*),topsep=0pt,itemsep=6pt,parsep=0pt,partopsep=0pt]
\item {\color{black}Fix a sequence of score descriptors $(\vartheta_N)_{N\ge1}$
and levels $\tau,\alpha\in(0,1)$.
For $P_X$-almost every $x$, there exist positive constants
$C_{x,\tau}$ and $C_{x,\alpha}$, independent of $N$ and $h_N$,
such that, for every $N$ and $0<h_N\le h_0$,
\[
 \mathcal L^{\mathrm{dep}}_{P,N,\tau}(\widehat q^{\mathrm{ker}}\mid x)
 \le C_{x,\tau}\left\{h_N^\zeta+\sqrt{\frac{h_N^{-d}+D_N}{n_N}}\right\}.
\]
}%
The response sets from TRACK and its variants satisfy
{\color{black}
\[
 \left|P\{Y_{N+1:N+H_N}\in\widehat C_N(x)\mid X_{N+1}=x\}-(1-\alpha)\right|
 \le C_{x,\alpha}\left\{h_N^\zeta+\sqrt{\frac{h_N^{-d}+D_N}{n_N}}\right\}.
\]
}%
Here $H_N=1$ for one-step prediction.
\item Fix $\tau_0\in(0,1/2)$. There exists $C>0$,
independent of $N,h_N,\vartheta_N,\tau$, such that, for
$0<h_N\le h_0$ and $\tau\in[\tau_0,1-\tau_0]$,
\[
 \sup_{x\in\mathcal X}\mathcal L^{\mathrm{ind}}_{P,N,\tau}
 (\widehat q^{\mathrm{ker}}\mid x)
 \le C\left\{h_N^\zeta+\sqrt{\frac{h_N^{-d}+D_N}{n_N}}\right\},
\]
and
\[
 \sup_{x\in\mathcal X}\left|P\{R_{N+1}^{\circ}\in\widehat C_{N,\tau}(x)
 \mid X_{N+1}^{\circ}=x\}-\tau\right|
 \le C\left\{h_N^\zeta+\sqrt{\frac{h_N^{-d}+D_N}{n_N}}\right\}.
\]
\end{enumerate}
{\color{black}If $h_N\asymp n_N^{-1/(2\zeta+d)}$, the expressions in braces in
both (i) and (ii) can be replaced by
$n_N^{-\zeta/(2\zeta+d)}+\sqrt{D_N/n_N}$.}%
\end{theorem}
}
}%

{\color{black}\hypersetup{allcolors=black,citecolor=blue}%
{\color{black}\hypersetup{allcolors=black,citecolor=blue}
Theorem~\ref{thm:conditional-coverage} establishes pointwise conditional
coverage bounds for TRACK and its variants, with uniformity over the
covariate space in the independent benchmark. These guarantees do not
require Assumption~\ref{ass:local-density}.%
}%
}%

\begingroup
\color{black}\hypersetup{allcolors=black,citecolor=blue}
\subsection{Conditional validity for general predictors}
\label{subsec:general-conditional-validity}
We continue with the rolling-window approximable predictors in
Section~\ref{subsec:general-predictors}. To control conditional \textcolor{black}{coverage
discrepancy}, define the approximation failure probability given the future
covariate by
\[
 p_N(x):=P\!\left\{\max_{L_N\le t\le N}
 \left\|\widetilde g_{N,1,t}-g_{N,1}(\,\cdot\,;Z_t,\ldots,Z_{t-L_N+1})\right\|_\infty
 >\varepsilon_N\,\middle|\,X_{N+1}=x\right\}.
\]

\begin{proposition}[Conditional validity for general predictors]
\label{cor:conditional-approximation}
For TRACK and TRACK-Abs, consider rolling-window approximable
predictors $\widetilde g_{N,1,t}$ with an approximation $g_{N,1}$
satisfying Assumptions~\ref{ass:distribution}--\ref{ass:comparison-density}.
Take $h_N\asymp n_N^{-1/(2\zeta+d)}$.
The following bounds hold for either interval form in \eqref{eq:oracle-intervals}.
\begin{enumerate}[label=(\roman*),topsep=2pt,itemsep=6pt,parsep=0pt,partopsep=0pt]
\item {\color{black}Fix the sequence of score descriptors $(\vartheta_N)_{N\ge1}$
induced by $g_{N,1}$ and levels $\tau,\alpha\in(0,1)$.
For $P_X$-almost every $x$, there exist positive constants
$C_{x,\tau}$ and $C_{x,\alpha}$, independent of $N$ and $h_N$,
such that
{\small\medmuskip=2mu\thickmuskip=3mu
\[
 P\!\left\{\widetilde R_{N+1}\in\widetilde C^*_{N,\tau}(x)\mathbin\triangle\widetilde C_{N,\tau}(x)
 \mid X_{N+1}=x\right\}
 \le C_{x,\tau}\left\{n_N^{-\zeta/(2\zeta+d)}+\sqrt{\frac{D_N}{n_N}}+\varepsilon_N+p_N(x)\right\}.
\]
}%
}%
The response sets from TRACK and TRACK-Abs satisfy
{\color{black}\small\medmuskip=2mu\thickmuskip=3mu
\[
 \left|P\!\left\{Y_{N+1}\in\widetilde C_N(x)\mid X_{N+1}=x\right\}-(1-\alpha)\right|
 \le C_{x,\alpha}\left\{n_N^{-\zeta/(2\zeta+d)}+\sqrt{\frac{D_N}{n_N}}+\varepsilon_N+p_N(x)\right\}.
\]
}%
\item Fix $\tau\in(0,1)$. There exists $C_\tau>0$,
independent of $N,h_N,\vartheta_N$, such that
{\small\medmuskip=2mu\thickmuskip=3mu
\[
 \sup_x P\!\left\{\widetilde R_{N+1}^{\circ}\in\widetilde C^*_{N,\tau}(x)\mathbin\triangle\widetilde C_{N,\tau}(x)
 \mid X_{N+1}^{\circ}=x\right\}
 \le C_\tau\left\{n_N^{-\zeta/(2\zeta+d)}+\sqrt{\frac{D_N}{n_N}}+\varepsilon_N+\pi_N+\sup_xp_N(x)\right\},
\]
}%
and
{\small\medmuskip=2mu\thickmuskip=3mu
\[
 \sup_x\left|P\!\left\{\widetilde R_{N+1}^{\circ}\in\widetilde C_{N,\tau}(x)\mid X_{N+1}^{\circ}=x\right\}-\tau\right|
 \le C_\tau\left\{n_N^{-\zeta/(2\zeta+d)}+\sqrt{\frac{D_N}{n_N}}+\varepsilon_N+\pi_N+\sup_xp_N(x)\right\}.
\]
}
\end{enumerate}
\end{proposition}

For the actual future target, the bound involves $p_N(x)$ in place of
the unconditional failure probability $\pi_N$. The following example
bounds this term uniformly in $x$ and gives explicit rates for expanding OLS.

\begin{example}
\label{ex:direct-ols-conditional}
Consider the expanding OLS predictor in Example~\ref{ex:expanding-predictor}
for TRACK and TRACK-Abs. With $K(v)=\tfrac12\1\{|v|\le1\}$,
we have $d=\zeta=1$, $D_N=L_N$ and $n_N=N-L_N$.
For all sufficiently large $N$, the approximation bounds are
$\varepsilon_N=C\sqrt{\log N/L_N}$, $\pi_N=N^{-2}$ and
$\sup_{x\in[-1,1]}p_N(x)\le N^{-2}$.
With $h_N\asymp n_N^{-1/3}$, for every fixed $\alpha\in(0,1)$,
\begin{equation}\label{eq:ols-uniform-conditional}
 \sup_{x\in[-1,1]}\left|P\{Y_{N+1}\in\widetilde C_N(x)\mid X_{N+1}=x\}-(1-\alpha)\right|
 \le C_\alpha\left\{n_N^{-1/3}+\sqrt{\frac{L_N}{n_N}}+\sqrt{\frac{\log N}{L_N}}+N^{-2}\right\}.
\end{equation}
All bounds in Proposition~\ref{cor:conditional-approximation} have the same rate,
and those for the actual future target hold uniformly over $x\in[-1,1]$.
\end{example}
This example shows that estimation from all past observations can yield
uniform conditional coverage bounds with serially dependent errors.
The corresponding TRACK-J results and direct OLS example are given in
the Appendix.
\par
\endgroup

\endgroup%

\section[Minimax rates for score calibration]{Minimax rates for score calibration}
\label{sec:minimax}
\begingroup%
\color{black}%
\hypersetup{allcolors=black,citecolor=blue}%
The preceding sections establish finite-sample and pointwise guarantees
for score calibration. We now study the best achievable \textcolor{black}{rates}
from the observed raw sample $O_N=(Z_1,\ldots,Z_N)$.
{\color{black}In the i.i.d.\ setting, kernel conformal bands attain minimax
optimality under symmetric-difference loss {\color{blue}\citep{lei2014distribution}}.}
{\color{black}Minimax bounds for weighted coverage error have also been derived for
score quantile sets {\color{blue}\citep{areces2024fundamental}}. We instead study
disagreement with oracle sets for scores formed from overlapping
windows of a stationary dependent process. The known descriptor
$\vartheta_N=(a_N,b_N,\rho_N)$ may vary with $N$, allowing the forecasting
rule to change and the window lengths and horizon to grow.}
{\color{black}\hypersetup{allcolors=black,citecolor=blue}The minimax results concern the rolling-window score classes.}

\subsection{Optimal rates at a prescribed score span}
\label{subsec:prescribed-span}
We first specify the possible data-generating laws for a given known
descriptor $\vartheta_N$. {\color{black}Since} the descriptor determines the score,
the relevant class consists of {\color{black}raw process} laws whose induced
conditional score distributions satisfy the required regularity.

\begin{definition}[Law classes for a known descriptor family]
\label{def:minimax-class}
Fix $\zeta\in(0,1]$, $p>2$, $\nu\in\mathbb N$ \textcolor{black}{and} positive finite
constants $C_\beta,L_F,c_X,C_X,\kappa_X,r_X$.
Given a deterministic family of known descriptors
$\{\vartheta_N=(a_N,b_N,\rho_N)\}_N$, let
$\mathfrak P(\vartheta_N)$ be the class of {\color{black}raw process}
laws $P$ satisfying Assumption~{\color{black}\ref{ass:distribution}} and for which
the score induced by $\vartheta_N$ satisfies
Assumption~{\color{black}\ref{ass:conditional-score-cdf}} with these constants.
{\color{black}Write $\mathcal X_P=\operatorname{supp}(P_X)$.
All displayed constants} are common to the law classes
and independent of $N$.
\end{definition}

For \textcolor{black}{the score used by \textcolor{black}{TRACK}}, fixing the forecasting
rule and training length gives $(a_N,b_N)=(L_N,0)$, and the class
concerns the resulting \textcolor{black}{signed residual} distribution. For
\textcolor{black}{the score used by \mbox{\textcolor{black}{TRACK-J}}}, fixing the forecasting rule, rolling
scale length \textcolor{black}{and} horizon gives
$(a_N,b_N)=(M_N+L_N+H_N-1,H_N-1)$, and the class concerns the
normalized trajectory score.
Definition~{\color{black}\ref{def:minimax-class}} restricts $P$ and its induced
score distribution. It imposes no additional requirement on the
quantile estimator $\widehat q${\color{black}, and} we retain the full admissible class
$\mathcal D_N(\vartheta_N)$ in
{\color{black}\eqref{eq:admissible-quantile-class}}.

\Needspace{4\baselineskip}
For a nonempty
$\mathfrak P(\vartheta_N)$, write
\begin{equation}\label{eq:fixed-minimax-risk}
 \mathfrak R^\star_{N,\tau}(\vartheta_N)
 :=\inf_{\widehat q\in\mathcal D_N(\vartheta_N)}
   \sup_{P\in\mathfrak P(\vartheta_N)}
   \mathcal L^\star_{P,N,\tau}(\widehat q),
 \qquad \star\in\{\mathrm{ind},\mathrm{dep}\}.
\end{equation}%
We use the losses from {\color{black}\eqref{eq:probability-scale-loss}}
and~{\color{black}\eqref{eq:dependent-set-loss}}, with either interval form
in {\color{black}\eqref{eq:oracle-intervals}}.

{\color{black}\hypersetup{allcolors=black,citecolor=blue}%
We first ask how accurately scores can be calibrated when their
span is fixed at $D_N$, allowing the forecasting and scoring rule
to vary over all known maps with that span. The next definition
formalizes the smallest error that can be guaranteed uniformly
over these maps and their compatible data-generating laws.}

\Needspace{4\baselineskip}
\begin{definition}[\textcolor{black}{Minimax risk at a prescribed score span}]
\label{def:span-envelope}
Fix a deterministic integer sequence $0\le D_N<N$ and the constants
in Definition~{\color{black}\ref{def:minimax-class}}. Let $\Theta_N(D_N)$ contain
all descriptors $\vartheta=(a,b,\rho)$ with $a,b\in\mathbb N_0$,
$a+b=D_N$, measurable $\rho:\mathcal Z^{D_N+1}\to\mathbb R$
\textcolor{black}{and} $\mathfrak P(\vartheta)\ne\varnothing$.
A rule is a family
$\boldsymbol{\widehat q}_N=(\widehat q_{N,\vartheta})_{\vartheta\in\Theta_N(D_N)}$
in $\mathfrak D_N(D_N):=\prod_{\vartheta\in\Theta_N(D_N)}
\mathcal D_N(\vartheta)$. Define
\begin{equation}\label{eq:span-envelope-risks}
 {\color{black}\mathfrak M^\star_{N,\tau}(D_N)}
 :=\inf_{\boldsymbol{\widehat q}_N\in\mathfrak D_N(D_N)}
   \sup_{\vartheta\in\Theta_N(D_N)}
   \sup_{P\in\mathfrak P(\vartheta)}
   \mathcal L^\star_{P,N,\tau}(\widehat q_{N,\vartheta}),
 \qquad \star\in\{\mathrm{ind},\mathrm{dep}\}.
\end{equation}%
\end{definition}

{\color{black}\hypersetup{allcolors=black,citecolor=blue}%
In practice, detailed properties of a fitted forecasting rule are
often unavailable. We therefore impose no smoothness or structural
restriction on $\rho$ beyond measurability \textcolor{black}{and} allow $\widehat q$
to be any member of $\mathcal D_N(\vartheta)$, including a randomized
rule, without additional restrictions.}
Theorem~{\color{black}\ref{thm:minimax-frontier}} first considers this broad class
of procedures and gives the optimal uniform rate at span $D_N$.

\Needspace{4\baselineskip}
\begin{theorem}[Minimax rate at a prescribed score span]
\label{thm:minimax-frontier}
{\color{black}Fix $\tau_0\in(0,1/2)$ and assume
$0<\kappa_X\le\Gamma(1+d/2)/(32\sqrt{\pi d})^d$.}
There is $N_0<\infty$ such that, for all $N\ge N_0$, integers
{\color{black}$0\le D_N<N$} \textcolor{black}{and}
$\tau\in[\tau_0,1-\tau_0]$, either interval form in
{\color{black}\eqref{eq:oracle-intervals}} satisfies
\begin{align}
 {\color{black}\mathfrak M^{\mathrm{ind}}_{N,\tau}(D_N)}
 &\asymp {\color{black}1\wedge
 \left\{n_N^{-\zeta/(2\zeta+d)}+\sqrt{D_N/n_N}\right\}},
 \label{eq:minimax-independent-frontier-rate}\\
 {\color{black}\mathfrak M^{\mathrm{dep}}_{N,\tau}(D_N)}
 &\asymp {\color{black}1\wedge
 \left\{n_N^{-\zeta/(2\zeta+d)}+\sqrt{D_N/n_N}\right\}}.
 \label{eq:minimax-frontier}
\end{align}%
The independent rate also holds when the loss in
{\color{black}\eqref{eq:span-envelope-risks}} is replaced by
$\sup_{x\in\mathcal X_P}\mathcal L^{\mathrm{ind}}_{P,N,\tau}
(\widehat q\mid x)$.
The positive finite constants implicit in $\asymp$ are independent
of $N$, $D_N$ \textcolor{black}{and} $\tau$.
\end{theorem}

The first term describes the difficulty of learning how the score
distribution changes with the covariate. The second describes the
largest additional error caused by overlap among score windows of
the prescribed span. The conclusion also holds when $D_N$
is an upper bound on the span in Definition~{\color{black}\ref{def:span-envelope}}.

Theorem~{\color{black}\ref{thm:minimax-frontier}} ranges over all known maps at
the same span. \textcolor{black}{For a given descriptor $\vartheta_N$, however, the score map $\rho_N$ is held fixed in the minimax risk.} This raises a further
question: does fixing $\rho_N$ leave the minimax rate unchanged,
or can maps with the same span have different calibration difficulty?

\Needspace{4\baselineskip}
\subsection{Why span alone does not determine the rate}
\label{subsec:fixed-map-rates}
We now consider calibration after the forecasting and scoring rule
has been specified. For a given deterministic descriptor family
$\{\vartheta_N\}_N$, the relevant quantity is
$\mathfrak R^\star_{N,\tau}(\vartheta_N)$ from
{\color{black}\eqref{eq:fixed-minimax-risk}}: the map is held fixed, while the risk
is minimized over all admissible quantile estimators and maximized
over the compatible laws $\mathfrak P(\vartheta_N)$. This is the
calibration difficulty associated with the chosen rule.

The span describes the extent of a score window \textcolor{black}{but} does not fully
describe how observations within that window enter the score.
The following proposition exhibits deterministic descriptor families
with span exactly $D_N$ whose minimax rates vary with the choice of
map, under the same \textcolor{black}{law class} constants.

\Needspace{4\baselineskip}
\begin{proposition}[\textcolor{black}{Minimax rates for score maps with a common span}]
\label{prop:same-span-different-rates}
{\color{black}\hypersetup{allcolors=black,citecolor=blue}%
{\color{black}Fix $\tau_0\in(0,1/2)$ and assume
$0<\kappa_X\le\Gamma(1+d/2)/(32\sqrt{\pi d})^d$.}
There are $c_0\in(0,1)$ and $N_0<\infty$ such that,
{\color{black}for every deterministic sequence of integer pairs
$\{(s_N,D_N)\}_N$} satisfying $1\le s_N\le D_N\le c_0N$,
there exists a deterministic family of known descriptors
$\{\vartheta_N^{(s_N)}\}_N$, each with span exactly $D_N$, satisfying}
\begin{equation}\label{eq:active-memory-frontier}
 \mathfrak R^\star_{N,\tau}(\vartheta_N^{(s_N)})
 \asymp {\color{black}n_N^{-\zeta/(2\zeta+d)}+\sqrt{s_N/n_N}},
 \qquad \star\in\{\mathrm{ind},\mathrm{dep}\},
\end{equation}%
for $N\ge N_0$ and $\tau\in[\tau_0,1-\tau_0]$, for either interval
form. The independent rate also holds when the loss
in {\color{black}\eqref{eq:fixed-minimax-risk}} is replaced by
$\sup_{x\in\mathcal X_P}\mathcal L^{\mathrm{ind}}_{P,N,\tau}
(\widehat q\mid x)$. The positive finite
{\color{black}constants implicit in $\asymp$} are
independent of $N,D_N,s_N,\tau$.
\end{proposition}

\begin{remark}
\label{rem:same-span-regimes}
The following choices of $s_N$ give the corresponding minimax rates
in {\color{black}\eqref{eq:active-memory-frontier}}:
\begin{enumerate}[label=(\roman*),topsep=4pt,itemsep=3pt,parsep=0pt]
 \item If $s_N=D_N$,
 $\mathfrak R^\star_{N,\tau}(\vartheta_N^{(D_N)})
 \asymp {\color{black}n_N^{-\zeta/(2\zeta+d)}+\sqrt{D_N/n_N}}$.
 \item If $s_N=1$,
 $\mathfrak R^\star_{N,\tau}(\vartheta_N^{(1)})
 \asymp {\color{black}n_N^{-\zeta/(2\zeta+d)}}$.
 \item If $s_N=\lfloor D_N/2\rfloor\vee1$
 {\color{black}and $D_N/n_N^{d/(2\zeta+d)}\to\infty$},
 $\mathfrak R^\star_{N,\tau}(\vartheta_N^{(s_N)})
 \asymp {\color{black}\sqrt{D_N/n_N}}$.
 \item {\color{black}If $0<\kappa<\zeta/(2\zeta+d)$,
 $D_N=\lfloor c_0N/2\rfloor$ \textcolor{black}{and}
 $s_N=\lceil n_N^{1-2\kappa}\rceil$},
 $\mathfrak R^\star_{N,\tau}(\vartheta_N^{(s_N)})
 \asymp {\color{black}n_N^{-\kappa}}$.
\end{enumerate}

These examples show that different forecasting and scoring rules
can have different minimax rates even when their spans agree.
Theorem~{\color{black}\ref{thm:minimax-frontier}}
gives a sharp rate uniformly over all known maps of a prescribed
span. For a fixed map, this rate remains an upper bound, but
Proposition~{\color{black}\ref{prop:same-span-different-rates}} shows that it need
not be sharp. Thus specifying $D_N$ alone does not determine the
minimax rate for a given forecasting and scoring rule. Further
information about how the map uses the observations is needed
to characterize its calibration difficulty.
\end{remark}
\endgroup%

\section{Numerical experiments}\label{sec:num}
\begingroup
\color{black}\hypersetup{allcolors=black,citecolor=blue}
\setlength{\intextsep}{6pt plus 1pt minus 1pt}
\clubpenalty=10000
\widowpenalty=10000
\captionsetup[table]{skip=4pt,font=small}
\newcommand{\best}[1]{\textbf{#1}}
\newcommand{\secondbest}[1]{\textbf{#1}}
\newcommand{\thirdbest}[1]{\textbf{#1}}
\newcommand{\metric}[2]{#1\,{\fontsize{7}{8}\selectfont(#2)}}
\newsavebox{\numericaltablebox}
\newcommand{\fitnumtable}[1]{%
  \sbox{\numericaltablebox}{#1}%
  \ifdim\wd\numericaltablebox>\textwidth
    \resizebox{\textwidth}{!}{\usebox{\numericaltablebox}}%
  \else\usebox{\numericaltablebox}\fi}

\textcolor{black}{We conduct numerical experiments to assess the calibration
and efficiency of \textcolor{black}{\textcolor{black}{TRACK} and its variants}} at
nominal coverage $0.9$ and horizons $H=1,3,10$. Local calibration uses
a product Epanechnikov kernel with $h_n\propto n^{-1/(d+2)}$, where
$n$ is the calibration history length and $d$ is the localization
dimension. Implementation details are given in
the Appendix.

{\color{black}%
{\color{black}\hypersetup{allcolors=black,citecolor=blue}%
Our experiments use covariates of dimension one or two for local calibration,
\textcolor{black}{with three and four evaluation regions respectively.} In the one-dimensional
case, we divide the observed evaluation range into three intervals of equal
width. Regions~1--3 correspond to low, middle \textcolor{black}{and} high covariate values.
In the two-dimensional case, we bisect each coordinate range into a lower
and an upper half. Their combinations form a $2\times2$ grid, with
regions~1--4 ordered low/low, high/low, low/high \textcolor{black}{and} high/high.
These partitions are used only for evaluation \textcolor{black}{as intended}.
}%

Global coverage is computed over all evaluation points, whereas regional
coverage is computed within each region. Both measure the fraction of
complete response vectors contained in their prediction sets, requiring
simultaneous coverage of all $H$ responses when $H>1$. If $p_{r,b}$ denotes
coverage in region $b$ of replication $r$, we report
\[
 \mathrm{R\text{-}MACE}=\frac1{MB}\sum_{r=1}^M\sum_{b=1}^B|p_{r,b}-0.9|,
\]
where $M$ is the number of replications and $B\in\{3,4\}$ is the number
of evaluation regions.%
}
Mean and Median summarize interval width ($H=1$) or root volume
$V^{1/H}$ ($H>1$), computed within replications and then averaged.
Parentheses give sample standard deviations across replications.

{\color{black}%
Within each setting and horizon, boldface indicates the top three of the
eleven methods for $H=1$ and the best of the two methods for $H>1$.
\textcolor{black}{Rankings are determined separately for each column: values closest to
$0.9$ are preferred in the Global and Region columns, and smaller values
in R-MACE. For Mean and Median, smaller interval widths or root volumes
are preferred among methods with Global values in \textcolor{black}{the desirable range of} $[0.88,0.92]$.}
}

\Needspace{8\baselineskip}
\subsection{Simulation studies}\label{sec:simulation}
{\color{black}\hypersetup{allcolors=black,citecolor=blue}%
For one-step-ahead prediction, we compare \textcolor{black}{TRACK and \mbox{TRACK-Abs}} with nine benchmark methods. {\color{black}\hypersetup{allcolors=black,citecolor=blue}Split conformal
prediction (SCP; {\color{blue}\citealp{lei2018distribution}}) provides a baseline based
on empirical residual quantiles, implemented here with an expanding
history of past forecast errors.}
The ACI family adapts calibration in response to past coverage errors.
ACI {\color{blue}\citep{gibbs2021adaptive}} updates the miscoverage level using a fixed
learning rate, while AgACI {\color{blue}\citep{zaffran2022adaptive}} and DtACI
{\color{blue}\citep{gibbs2024conformal}} extend this approach by combining ACI experts
with different learning rates. SF-OGD {\color{blue}\citep{bhatnagar2023improved}} also
uses coverage feedback to update the score threshold through
scale-free online gradient descent. PI and PID {\color{blue}\citep{angelopoulos2023pid}}
take a related approach based on feedback control: PI combines quantile
tracking with correction for accumulated coverage errors, while PID
additionally incorporates forecasts of future scores through a scorecaster.
{\color{black}\hypersetup{allcolors=black,citecolor=blue}EnbPI {\color{blue}\citep{xu2023conformal}} and SPCI
{\color{blue}\citep{xu2023sequential}} combine bootstrap ensemble forecasts with
residual calibration.} EnbPI uses empirical residual quantiles, whereas
SPCI estimates residual quantiles conditional on recent prediction
errors to account for temporal dependence.

At $H=3$ and $H=10$, we compare \textcolor{black}{\mbox{TRACK-J}} with the unweighted Bonferroni
version of the conformal multi-step method (Conformal-MS)
{\color{blue}\hypersetup{citecolor=blue}\citep{schlembach2022conformal}}. Comparator configurations are given
in the Appendix.
}

{\color{black}\hypersetup{allcolors=black,citecolor=blue}%
{\color{black}\hypersetup{allcolors=black,citecolor=blue}%
Simulation studies of conformal prediction for time series have used
relatively large sample sizes, such as $2000$ {\color{blue}\citep{xu2023sequential}}
or $6000$ observations {\color{blue}\citep{gibbs2024conformal}}.
\textcolor{black}{We choose $T=5000$ observations and use the last $1000$ time points as test points to evaluate all performance metrics at each horizon. This provides enough test points within each covariate region while retaining sufficient calibration history.}
\textcolor{black}{Experiments with $T=2000$ and $T=10000$ show similar performance, with detailed results provided in the Appendix.}
Each setting uses $M=100$ independent replications.
}%
Point forecasts use ordinary least squares (OLS) without an intercept
in Settings~1 to~3 and a random forest with $100$ trees in Setting~4,
refitted on a rolling window of length $L=100$.
EnbPI and SPCI instead retain their native bootstrap forecasts with
the corresponding model class and covariates.
}%

\subsubsection{\textcolor{black}{Data generating processes}}\label{sec:simulation-factorial}
We consider the following four settings of the form
\[
 Y_t=m(X_t)+\sigma_t\epsilon_t.
\]
Settings~1 to~3 have i.i.d. $X_t\sim\operatorname{Unif}([0,1]^2)$,
independent of the i.i.d. errors, with $m(x)=x_1-2x_2$ and
$\sigma_t=\sigma(X_t)$.

\paragraph{Setting 1: homoscedastic Gaussian.}
The scale is $\sigma(x)=1$ and $\epsilon_t\sim\mathcal N(0,1)$.

\paragraph{Setting 2: Gaussian with interaction scale.}
The errors remain standard Gaussian, while
$\sigma(x)=0.5+3x_1x_2$ makes the variance depend on both covariates.

\paragraph{Setting 3: CNL with interaction scale.}
We retain $\sigma(x)=0.5+3x_1x_2$ and set
$\epsilon_t=U_t-\mu_U$. The connected normal--Laplace (CNL) law
{\color{blue}\citep{zhao2026minimax}} has a normal branch on the negative half line
and a Laplace branch on the positive half line, each with probability $1/2$:
\[
 f_U(u)=\frac1{\sqrt{2\pi}}\begin{cases}
 e^{-u^2/2},&u\le0,\\
 e^{-\sqrt{2/\pi}\,u},&u>0,
 \end{cases}
 \qquad \mu_U=\E U=\frac{\pi-2}{2\sqrt{2\pi}}.
\]

\paragraph{Setting 4: nonlinear ARMA--GARCH.}
Take independent i.i.d. $\xi_t,\epsilon_t\sim\mathcal N(0,1)$ and
$\sigma_t=\sqrt{v_t}$, with
\begin{equation*}
 \begin{aligned}
 m(x)&=e^{x^2/4}+\log(1+x^2)+5|x|+x^2+5\sin(2\pi x)+10\cos(\pi x),\\
 X_t&=0.6X_{t-1}+\xi_t+0.2\xi_{t-1},\\
 v_t&=0.2+0.15v_{t-1}\epsilon_{t-1}^2+0.6v_{t-1}.
 \end{aligned}
\end{equation*}

{\color{black}\emergencystretch=1em
The covariate process is an autoregressive moving average model
ARMA(1,1). The conditional error variance follows a generalized
autoregressive conditional heteroscedasticity model GARCH(1,1).\par}

Setting~1 provides a benchmark with constant uncertainty. Settings~2
and~3 introduce variance differences across covariate regions under
the same regression mean, with the CNL law adding asymmetric tails.
Setting~4 combines nonlinear regression, temporal dependence \textcolor{black}{and}
persistent volatility. These designs assess whether local calibration
improves regional coverage as the prediction problem becomes more complex.

\subsubsection{\textcolor{black}{One-step-ahead prediction}}\label{sec:simulation-one-step}
{\color{black}\hypersetup{allcolors=black,citecolor=blue}%
We first consider one-step-ahead prediction.
Tables~{\color{black}\ref{tab:simulation-one-selected}} and~{\color{black}\ref{tab:simulation-one-complex}}
present the Gaussian designs and the more complex designs \textcolor{black}{respectively}.
The Gaussian settings use simple symmetric errors, whereas the remaining
settings introduce asymmetric errors or combine nonlinear regression
with temporal dependence and changing volatility.
}%

\begin{table}[!htbp]
\color{black}\hypersetup{allcolors=black,citecolor=blue}%
\centering\setstretch{1}\fontsize{9}{10.5}\selectfont
\setlength{\tabcolsep}{1.5pt}
\renewcommand{\arraystretch}{1.03}
\fitnumtable{\begin{tabular}{@{}l*{8}{c}@{}}
\toprule
Method & Global & Mean & Median & Region~1 & Region~2 & Region~3 & Region~4 & R-MACE \\
\midrule
\multicolumn{9}{@{}l}{\textit{Setting~1: Gaussian, homoscedastic}}\\*
TRACK & \metric{0.894}{0.01} & \secondbest{\metric{3.320}{0.05}} & \secondbest{\metric{3.312}{0.05}} & \metric{0.894}{0.02} & \metric{0.895}{0.02} & \metric{0.894}{0.02} & \metric{0.894}{0.02} & 0.016 \\*
\mbox{TRACK-Abs} & \metric{0.897}{0.01} & \thirdbest{\metric{3.325}{0.05}} & \thirdbest{\metric{3.316}{0.05}} & \metric{0.897}{0.02} & \metric{0.897}{0.02} & \metric{0.899}{0.02} & \metric{0.897}{0.02} & 0.015 \\*
ACI & \best{\metric{0.900}{0.00}} & \metric{3.341}{0.09} & \metric{3.327}{0.09} & \metric{0.902}{0.02} & \metric{0.899}{0.02} & \thirdbest{\metric{0.899}{0.02}} & \metric{0.899}{0.02} & \best{0.014} \\*
AgACI & \metric{0.901}{0.01} & \metric{3.338}{0.08} & \metric{3.334}{0.08} & \metric{0.904}{0.02} & \secondbest{\metric{0.900}{0.02}} & \secondbest{\metric{0.900}{0.02}} & \secondbest{\metric{0.900}{0.02}} & \thirdbest{0.014} \\*
DtACI & \metric{0.899}{0.00} & \metric{3.330}{0.08} & \metric{3.330}{0.08} & \metric{0.902}{0.02} & \metric{0.898}{0.02} & \metric{0.899}{0.02} & \metric{0.898}{0.02} & \secondbest{0.014} \\*
SF-OGD & \metric{0.617}{0.02} & \metric{1.763}{0.01} & \metric{1.765}{0.01} & \metric{0.620}{0.03} & \metric{0.616}{0.03} & \metric{0.618}{0.03} & \metric{0.614}{0.03} & 0.283 \\*
SCP & \thirdbest{\metric{0.900}{0.01}} & \metric{3.329}{0.05} & \metric{3.329}{0.05} & \metric{0.903}{0.02} & \thirdbest{\metric{0.900}{0.02}} & \best{\metric{0.900}{0.02}} & \thirdbest{\metric{0.899}{0.02}} & 0.015 \\*
PI & \metric{0.899}{0.00} & \metric{3.514}{0.08} & \metric{3.502}{0.09} & \best{\metric{0.900}{0.02}} & \metric{0.898}{0.02} & \metric{0.898}{0.02} & \metric{0.899}{0.02} & 0.015 \\*
PID & \metric{0.899}{0.00} & \metric{3.527}{0.09} & \metric{3.514}{0.09} & \thirdbest{\metric{0.902}{0.02}} & \metric{0.898}{0.02} & \metric{0.899}{0.02} & \metric{0.896}{0.02} & 0.014 \\*
EnbPI & \secondbest{\metric{0.900}{0.01}} & \best{\metric{3.295}{0.05}} & \best{\metric{3.295}{0.05}} & \secondbest{\metric{0.901}{0.02}} & \best{\metric{0.900}{0.02}} & \metric{0.901}{0.02} & \best{\metric{0.900}{0.02}} & 0.015 \\*
SPCI & \metric{0.860}{0.01} & \metric{3.043}{0.04} & \metric{3.262}{0.05} & \metric{0.862}{0.02} & \metric{0.860}{0.02} & \metric{0.863}{0.02} & \metric{0.857}{0.02} & 0.040 \\
\midrule
\multicolumn{9}{@{}l}{\textit{Setting~2: Gaussian, interaction}}\\*
TRACK & \metric{0.898}{0.01} & \best{\metric{4.191}{0.08}} & \best{\metric{3.564}{0.11}} & \secondbest{\metric{0.899}{0.02}} & \best{\metric{0.900}{0.02}} & \best{\metric{0.900}{0.02}} & \secondbest{\metric{0.893}{0.02}} & \secondbest{0.015} \\*
\mbox{TRACK-Abs} & \metric{0.901}{0.01} & \secondbest{\metric{4.198}{0.09}} & \secondbest{\metric{3.571}{0.11}} & \best{\metric{0.901}{0.02}} & \thirdbest{\metric{0.904}{0.02}} & \secondbest{\metric{0.904}{0.02}} & \best{\metric{0.896}{0.02}} & \best{0.015} \\*
ACI & \best{\metric{0.900}{0.00}} & \metric{4.652}{0.21} & \metric{4.614}{0.21} & \metric{0.995}{0.00} & \metric{0.946}{0.01} & \metric{0.947}{0.01} & \metric{0.714}{0.02} & 0.094 \\*
AgACI & \secondbest{\metric{0.900}{0.00}} & \metric{4.632}{0.19} & \metric{4.620}{0.19} & \metric{0.995}{0.00} & \metric{0.948}{0.01} & \metric{0.948}{0.01} & \metric{0.713}{0.02} & 0.095 \\*
DtACI & \thirdbest{\metric{0.899}{0.00}} & \metric{4.622}{0.18} & \metric{4.611}{0.19} & \metric{0.995}{0.00} & \metric{0.946}{0.01} & \metric{0.947}{0.02} & \metric{0.710}{0.02} & 0.094 \\*
SF-OGD & \metric{0.590}{0.02} & \metric{1.790}{0.01} & \metric{1.792}{0.01} & \metric{0.802}{0.02} & \metric{0.617}{0.03} & \metric{0.616}{0.03} & \metric{0.328}{0.03} & 0.309 \\*
SCP & \metric{0.898}{0.01} & \metric{4.583}{0.09} & \metric{4.585}{0.09} & \metric{0.995}{0.00} & \metric{0.946}{0.01} & \metric{0.947}{0.02} & \metric{0.706}{0.03} & 0.095 \\*
PI & \metric{0.898}{0.00} & \metric{5.050}{0.21} & \metric{5.009}{0.23} & \metric{0.989}{0.01} & \metric{0.938}{0.01} & \metric{0.939}{0.01} & \metric{0.729}{0.02} & \thirdbest{0.084} \\*
PID & \metric{0.898}{0.00} & \metric{5.079}{0.21} & \metric{5.034}{0.22} & \metric{0.989}{0.01} & \metric{0.939}{0.01} & \metric{0.938}{0.01} & \thirdbest{\metric{0.729}{0.02}} & 0.084 \\*
EnbPI & \metric{0.898}{0.01} & \thirdbest{\metric{4.527}{0.09}} & \thirdbest{\metric{4.527}{0.09}} & \metric{0.995}{0.00} & \metric{0.946}{0.01} & \metric{0.946}{0.02} & \metric{0.705}{0.03} & 0.096 \\*
SPCI & \metric{0.857}{0.01} & \metric{4.083}{0.08} & \metric{4.457}{0.09} & \thirdbest{\metric{0.975}{0.01}} & \secondbest{\metric{0.903}{0.02}} & \thirdbest{\metric{0.905}{0.02}} & \metric{0.648}{0.03} & 0.090 \\
\bottomrule
\end{tabular}}
\caption{\textcolor{black}{One-step-ahead prediction: Gaussian designs.}}
\label{tab:simulation-one-selected}
\end{table}

{\color{black}\hypersetup{allcolors=black,citecolor=blue}%
In the homoscedastic benchmark, most methods attain satisfactory
marginal and regional coverage with comparable interval widths.
TRACK and \mbox{TRACK-Abs} remain competitive in this simple setting.

With interaction variance, most methods still attain marginal coverage
close to the nominal level, but their regional coverage differs
substantially. TRACK and \mbox{TRACK-Abs} outperform all competitors in
regional calibration and produce shorter intervals than competing
methods meeting the global coverage criterion. This improvement
reflects the benefit of adapting to the local error distribution when
marginal coverage alone conceals systematic differences across regions.
}%

\begin{table}[!htbp]
\color{black}\hypersetup{allcolors=black,citecolor=blue}%
\centering\setstretch{1}\fontsize{9}{10.5}\selectfont
\setlength{\tabcolsep}{1.5pt}
\renewcommand{\arraystretch}{1.03}
\fitnumtable{\begin{tabular}{@{}l*{8}{c}@{}}
\toprule
Method & Global & Mean & Median & Region~1 & Region~2 & Region~3 & Region~4 & R-MACE \\
\midrule
\multicolumn{9}{@{}l}{\textit{Setting~3: CNL, interaction}}\\*
TRACK & \metric{0.897}{0.01} & \secondbest{\metric{5.785}{0.15}} & \secondbest{\metric{4.907}{0.19}} & \best{\metric{0.899}{0.02}} & \secondbest{\metric{0.896}{0.02}} & \best{\metric{0.899}{0.02}} & \secondbest{\metric{0.895}{0.02}} & \secondbest{0.016} \\*
\mbox{TRACK-Abs} & \secondbest{\metric{0.900}{0.01}} & \best{\metric{5.577}{0.12}} & \best{\metric{4.738}{0.18}} & \secondbest{\metric{0.901}{0.02}} & \best{\metric{0.900}{0.02}} & \secondbest{\metric{0.902}{0.02}} & \best{\metric{0.898}{0.02}} & \best{0.015} \\*
ACI & \thirdbest{\metric{0.900}{0.00}} & \metric{6.102}{0.29} & \metric{6.051}{0.30} & \metric{0.985}{0.01} & \metric{0.937}{0.01} & \metric{0.939}{0.01} & \metric{0.739}{0.02} & 0.081 \\*
AgACI & \metric{0.901}{0.01} & \metric{6.099}{0.24} & \metric{6.069}{0.24} & \metric{0.986}{0.01} & \metric{0.939}{0.01} & \metric{0.939}{0.01} & \metric{0.741}{0.02} & 0.081 \\*
DtACI & \metric{0.899}{0.00} & \metric{6.059}{0.24} & \metric{6.044}{0.25} & \metric{0.985}{0.01} & \metric{0.937}{0.01} & \metric{0.938}{0.01} & \metric{0.738}{0.02} & 0.080 \\*
SF-OGD & \metric{0.534}{0.02} & \metric{1.899}{0.01} & \metric{1.901}{0.01} & \metric{0.734}{0.03} & \metric{0.552}{0.03} & \metric{0.557}{0.03} & \metric{0.295}{0.03} & 0.365 \\*
SCP & \metric{0.899}{0.01} & \metric{6.021}{0.14} & \metric{6.021}{0.14} & \metric{0.985}{0.01} & \metric{0.938}{0.01} & \metric{0.939}{0.01} & \metric{0.738}{0.03} & 0.081 \\*
PI & \metric{0.898}{0.00} & \metric{6.847}{0.38} & \metric{6.755}{0.37} & \metric{0.982}{0.01} & \metric{0.932}{0.01} & \metric{0.934}{0.01} & \metric{0.746}{0.02} & 0.075 \\*
PID & \metric{0.898}{0.00} & \metric{6.866}{0.36} & \metric{6.775}{0.37} & \metric{0.981}{0.01} & \metric{0.931}{0.01} & \metric{0.935}{0.01} & \thirdbest{\metric{0.747}{0.02}} & \thirdbest{0.075} \\*
EnbPI & \best{\metric{0.900}{0.01}} & \thirdbest{\metric{5.941}{0.13}} & \thirdbest{\metric{5.941}{0.13}} & \metric{0.985}{0.01} & \metric{0.939}{0.01} & \metric{0.940}{0.01} & \metric{0.737}{0.02} & 0.082 \\*
SPCI & \metric{0.855}{0.01} & \metric{5.328}{0.12} & \metric{5.899}{0.13} & \thirdbest{\metric{0.958}{0.01}} & \thirdbest{\metric{0.895}{0.02}} & \thirdbest{\metric{0.896}{0.02}} & \metric{0.674}{0.03} & 0.079 \\
\midrule
\multicolumn{9}{@{}l}{\textit{Setting~4: Nonlinear ARMA--GARCH}}\\*
TRACK & \metric{0.882}{0.01} & \secondbest{\metric{6.086}{0.42}} & \best{\metric{4.408}{0.12}} & \secondbest{\metric{0.794}{0.06}} & \secondbest{\metric{0.893}{0.01}} & \secondbest{\metric{0.835}{0.05}} & --- & \secondbest{0.062} \\*
\mbox{TRACK-Abs} & \metric{0.893}{0.01} & \metric{7.810}{1.05} & \thirdbest{\metric{4.601}{0.14}} & \best{\metric{0.853}{0.06}} & \best{\metric{0.897}{0.01}} & \best{\metric{0.869}{0.04}} & --- & \best{0.034} \\*
ACI & \best{\metric{0.900}{0.00}} & \metric{6.441}{0.41} & \metric{6.264}{0.37} & \metric{0.726}{0.07} & \metric{0.950}{0.01} & \metric{0.555}{0.06} & --- & 0.190 \\*
AgACI & \metric{0.908}{0.01} & \metric{7.502}{1.87} & \metric{6.349}{0.32} & \metric{0.742}{0.07} & \metric{0.955}{0.01} & \metric{0.589}{0.06} & --- & 0.175 \\*
DtACI & \thirdbest{\metric{0.899}{0.00}} & \metric{6.527}{1.28} & \metric{6.228}{0.30} & \metric{0.725}{0.07} & \metric{0.948}{0.01} & \metric{0.559}{0.06} & --- & 0.188 \\*
SF-OGD & \metric{0.500}{0.02} & \metric{1.978}{0.01} & \metric{1.979}{0.01} & \metric{0.341}{0.07} & \metric{0.542}{0.02} & \metric{0.211}{0.04} & --- & 0.535 \\*
SCP & \secondbest{\metric{0.900}{0.01}} & \thirdbest{\metric{6.274}{0.17}} & \metric{6.275}{0.17} & \metric{0.725}{0.08} & \metric{0.951}{0.01} & \metric{0.547}{0.06} & --- & 0.193 \\*
PI & \metric{0.907}{0.01} & \metric{65.559}{226.48} & \metric{50.560}{220.41} & \thirdbest{\metric{0.764}{0.06}} & \metric{0.938}{0.01} & \thirdbest{\metric{0.704}{0.07}} & --- & \thirdbest{0.124} \\*
PID & \metric{0.907}{0.01} & \metric{95.581}{366.44} & \metric{80.282}{366.77} & \metric{0.764}{0.07} & \metric{0.939}{0.01} & \metric{0.702}{0.07} & --- & 0.125 \\*
EnbPI & \metric{0.901}{0.01} & \best{\metric{4.454}{0.20}} & \secondbest{\metric{4.456}{0.20}} & \metric{0.665}{0.08} & \metric{0.967}{0.01} & \metric{0.456}{0.07} & --- & 0.249 \\*
SPCI & \metric{0.857}{0.01} & \metric{5.640}{3.63} & \metric{4.378}{0.19} & \metric{0.626}{0.07} & \thirdbest{\metric{0.920}{0.01}} & \metric{0.439}{0.06} & --- & 0.252 \\
\bottomrule
\end{tabular}}
\caption{\textcolor{black}{One-step-ahead prediction: asymmetric and nonlinear designs.}}
\label{tab:simulation-one-complex}
\end{table}

{\color{black}\hypersetup{allcolors=black,citecolor=blue}%
Table~{\color{black}\ref{tab:simulation-one-complex}} extends the comparison to
asymmetric errors and nonlinear regression with temporal dependence.
Under the asymmetric CNL errors in Setting~3, TRACK and \mbox{TRACK-Abs}
maintain balanced regional coverage with short intervals. Setting~4
combines nonlinear regression, dependent covariates \textcolor{black}{and} persistent
volatility. Local calibration remains effective under this more complex
structure, although some undercoverage persists in \textcolor{black}{both tail regions}.
\mbox{TRACK-Abs} gives better regional balance. The unusually large
width summaries and standard deviations for PI and PID are driven by
a few replications with persistently wide intervals.
}%

\Needspace{9\baselineskip}
\subsubsection{\textcolor{black}{Multi-step-ahead prediction}}\label{sec:simulation-multi-step}
We next examine whether the regional benefits persist when all future
responses \textcolor{black}{over the specified forecast horizon} must be covered simultaneously. At origin $t$, the simulations
provide $X_{t+1}$ for forecasting $(Y_{t+1},\ldots,Y_{t+H})$.
Both methods share the point forecasts at each horizon and use fully
observed response vectors for calibration. \textcolor{black}{\mbox{TRACK-J}} forms ellipsoids
using the normalized score in {\color{black}\eqref{eq:normalized-score}}.
Conformal-MS forms rectangles using absolute residuals with a
Bonferroni correction. Table~{\color{black}\ref{tab:simulation-multi-selected}}
compares joint coverage and root volume at $H=3,10$ within each setting.

\begin{table}[!htbp]
\color{black}\hypersetup{allcolors=black,citecolor=blue}%
\centering\setstretch{1}\fontsize{9}{10.5}\selectfont
\setlength{\tabcolsep}{1.5pt}
\renewcommand{\arraystretch}{1.03}
\fitnumtable{\begin{tabular}{@{}l*{8}{c}@{}}
\toprule
Method & Global & Mean & Median & Region~1 & Region~2 & Region~3 & Region~4 & R-MACE \\
\midrule
\multicolumn{9}{@{}l}{\textit{$H=3$, Setting~1: Gaussian, homoscedastic}}\\*
\mbox{TRACK-J} & \best{\metric{0.898}{0.01}} & \best{\metric{4.796}{0.11}} & \best{\metric{4.770}{0.11}} & \metric{0.898}{0.02} & \best{\metric{0.899}{0.02}} & \best{\metric{0.897}{0.02}} & \metric{0.896}{0.02} & \best{0.015} \\
Conformal-MS & \metric{0.903}{0.01} & \metric{4.851}{0.08} & \metric{4.850}{0.08} & \best{\metric{0.899}{0.02}} & \metric{0.906}{0.02} & \metric{0.904}{0.02} & \best{\metric{0.900}{0.02}} & 0.018 \\
\midrule
\multicolumn{9}{@{}l}{\textit{$H=10$, Setting~1: Gaussian, homoscedastic}}\\*
\mbox{TRACK-J} & \best{\metric{0.899}{0.01}} & \best{\metric{5.583}{0.14}} & \best{\metric{5.559}{0.14}} & \best{\metric{0.902}{0.02}} & \best{\metric{0.898}{0.02}} & \best{\metric{0.900}{0.02}} & \metric{0.898}{0.02} & \best{0.017} \\
Conformal-MS & \metric{0.902}{0.02} & \metric{6.085}{0.12} & \metric{6.085}{0.12} & \metric{0.897}{0.03} & \metric{0.906}{0.03} & \metric{0.906}{0.03} & \best{\metric{0.901}{0.03}} & 0.024 \\
\midrule
\multicolumn{9}{@{}l}{\textit{$H=3$, Setting~2: Gaussian, interaction}}\\*
\mbox{TRACK-J} & \metric{0.897}{0.01} & \best{\metric{6.812}{0.23}} & \best{\metric{6.519}{0.23}} & \best{\metric{0.898}{0.02}} & \best{\metric{0.897}{0.02}} & \best{\metric{0.898}{0.02}} & \best{\metric{0.896}{0.02}} & \best{0.014} \\
Conformal-MS & \best{\metric{0.902}{0.02}} & \metric{7.184}{0.22} & \metric{7.185}{0.22} & \metric{0.931}{0.02} & \metric{0.926}{0.02} & \metric{0.926}{0.02} & \metric{0.824}{0.03} & 0.042 \\
\midrule
\multicolumn{9}{@{}l}{\textit{$H=10$, Setting~2: Gaussian, interaction}}\\*
\mbox{TRACK-J} & \best{\metric{0.899}{0.02}} & \best{\metric{8.061}{0.28}} & \best{\metric{7.903}{0.27}} & \best{\metric{0.901}{0.02}} & \best{\metric{0.899}{0.02}} & \best{\metric{0.901}{0.02}} & \best{\metric{0.897}{0.03}} & \best{0.018} \\
Conformal-MS & \metric{0.904}{0.03} & \metric{9.855}{0.37} & \metric{9.857}{0.38} & \metric{0.911}{0.03} & \metric{0.911}{0.03} & \metric{0.912}{0.03} & \metric{0.884}{0.03} & 0.028 \\
\midrule
\multicolumn{9}{@{}l}{\textit{$H=3$, Setting~3: CNL, interaction}}\\*
\mbox{TRACK-J} & \best{\metric{0.898}{0.01}} & \best{\metric{9.620}{0.52}} & \best{\metric{8.946}{0.43}} & \best{\metric{0.897}{0.02}} & \best{\metric{0.898}{0.02}} & \best{\metric{0.900}{0.02}} & \best{\metric{0.898}{0.02}} & \best{0.016} \\
Conformal-MS & \metric{0.903}{0.02} & \metric{9.875}{0.32} & \metric{9.874}{0.32} & \metric{0.933}{0.02} & \metric{0.919}{0.02} & \metric{0.921}{0.02} & \metric{0.840}{0.03} & 0.035 \\
\midrule
\multicolumn{9}{@{}l}{\textit{$H=10$, Setting~3: CNL, interaction}}\\*
\mbox{TRACK-J} & \best{\metric{0.901}{0.02}} & \best{\metric{11.716}{0.65}} & \best{\metric{11.130}{0.61}} & \best{\metric{0.900}{0.02}} & \best{\metric{0.902}{0.02}} & \best{\metric{0.900}{0.02}} & \best{\metric{0.900}{0.02}} & \best{0.019} \\
Conformal-MS & \metric{0.906}{0.03} & \metric{14.645}{0.76} & \metric{14.631}{0.77} & \metric{0.917}{0.03} & \metric{0.913}{0.03} & \metric{0.910}{0.03} & \metric{0.884}{0.03} & 0.027 \\
\midrule
\multicolumn{9}{@{}l}{\textit{$H=3$, Setting~4: Nonlinear ARMA--GARCH}}\\*
\mbox{TRACK-J} & \best{\metric{0.892}{0.01}} & \metric{112.564}{229.90} & \best{\metric{31.609}{2.86}} & \best{\metric{0.849}{0.07}} & \best{\metric{0.897}{0.01}} & \best{\metric{0.872}{0.04}} & --- & \best{0.038} \\
Conformal-MS & \metric{0.916}{0.02} & \best{\metric{38.745}{3.73}} & \metric{38.744}{3.70} & \metric{0.676}{0.10} & \metric{0.956}{0.01} & \metric{0.691}{0.07} & --- & 0.164 \\
\midrule
\multicolumn{9}{@{}l}{\textit{$H=10$, Setting~4: Nonlinear ARMA--GARCH}}\\*
\mbox{TRACK-J} & \best{\metric{0.890}{0.02}} & \metric{161.091}{273.24} & \best{\metric{68.689}{7.32}} & \best{\metric{0.848}{0.06}} & \best{\metric{0.894}{0.02}} & \best{\metric{0.874}{0.05}} & --- & \best{0.040} \\
Conformal-MS & \metric{0.920}{0.03} & \best{\metric{140.999}{34.24}} & \metric{140.629}{33.46} & \metric{0.769}{0.10} & \metric{0.936}{0.03} & \metric{0.862}{0.07} & --- & 0.079 \\
\bottomrule
\end{tabular}}
\caption{\textcolor{black}{Multi-step-ahead prediction: simulation studies.}}
\label{tab:simulation-multi-selected}
\end{table}

{\color{black}\hypersetup{allcolors=black,citecolor=blue}\emergencystretch=3em%
The results extend the effectiveness of local calibration to
multi-step-ahead prediction. \mbox{TRACK-J} maintains joint coverage close to
nominal and small regional calibration errors, with smaller median root
volumes across all four settings. These coverage and efficiency gains
persist at $H=10$, supporting the method's effectiveness over a longer
forecasting horizon. The large mean \textcolor{black}{root volumes} in Setting~4
\textcolor{black}{at both horizons are}
driven by occasional extreme sets. With few nearby calibration observations, an extreme score can
dominate the weighted quantile and enlarge the prediction set. The median is less sensitive to
these extremes and therefore better reflects typical set size.\par
}%

\Needspace{6\baselineskip}
\subsection{Real data studies}\label{sec:real-data}
{\color{black}\hypersetup{allcolors=black,citecolor=blue}%
We study two forecasting applications with different response
characteristics. The wind data record hourly electricity generation
at the Hackberry Wind Project in Texas, a previously studied benchmark
{\color{blue}\citep{xu2023conformal}}. The original wind data contain $13871$
observations from January 2019 to July 2020, measured in MWh.
The bike data record hourly rentals from the Capital Bikeshare system
in Washington, D.C., together with weather and calendar information
{\color{blue}\citep{fanaee2014event}}. The original bike data contain $17379$
observations from 2011 and 2012.
}%

{\color{black}\hypersetup{allcolors=black,citecolor=blue}We consider the same eleven methods for one-step-ahead prediction and two methods for multi-step-ahead prediction as in Section~{\color{black}\ref{sec:simulation}}.}
Shared forecasts use random forests with $100$ trees and rolling window
$L=100$. \textcolor{black}{EnbPI and SPCI use the same random forest specification
within their own bootstrap ensembles.}
Each application uses $M=100$ runs on the same observed series,
evaluating the final $2000$ fully observed forecast origins at each
horizon. {\color{black}\hypersetup{allcolors=black,citecolor=blue}Further details are provided in the Appendix.}

\subsubsection{\textcolor{black}{One-step-ahead prediction}}\label{sec:real-one-step}
{\color{black}\hypersetup{allcolors=black,citecolor=blue}%
We assess whether the methods remain effective on real data,
where the data-generating distribution is unknown.
Table~{\color{black}\ref{tab:real-one-selected}} reports one-step-ahead coverage and
interval width for the two applications.
}%

\begin{table}[!htbp]
\centering\setstretch{1}\fontsize{9}{10.5}\selectfont
\setlength{\tabcolsep}{1.5pt}
\renewcommand{\arraystretch}{1.03}
\fitnumtable{\begin{tabular}{@{}l*{8}{c}@{}}
\toprule
Method & Global & Mean & Median & Region~1 & Region~2 & Region~3 & Region~4 & R-MACE \\
\midrule
\multicolumn{9}{@{}l}{\textit{Hackberry wind}}\\*
\textcolor{black}{TRACK} & \metric{0.894}{0.00} & \textcolor{black}{\best{\metric{62.535}{0.13}}} & \textcolor{black}{\best{\metric{60.762}{0.29}}} & \textcolor{black}{\best{\metric{0.907}{0.00}}} & \textcolor{black}{\thirdbest{\metric{0.836}{0.01}}} & \textcolor{black}{\metric{0.814}{0.01}} & \textcolor{black}{\metric{0.885}{0.01}} & \thirdbest{0.043} \\
\textcolor{black}{\mbox{TRACK-Abs}} & \metric{0.907}{0.00} & \textcolor{black}{\thirdbest{\metric{64.998}{0.14}}} & \textcolor{black}{\secondbest{\metric{62.154}{0.41}}} & \secondbest{\metric{0.913}{0.00}} & \best{\metric{0.896}{0.01}} & \textcolor{black}{\secondbest{\metric{0.891}{0.01}}} & \textcolor{black}{\best{\metric{0.895}{0.01}}} & \best{0.009} \\
ACI & \metric{0.905}{0.00} & \textcolor{black}{\metric{67.652}{0.58}} & \textcolor{black}{\metric{65.179}{0.80}} & \metric{0.934}{0.00} & \textcolor{black}{\metric{0.760}{0.01}} & \textcolor{black}{\metric{0.840}{0.01}} & \textcolor{black}{\metric{0.860}{0.00}} & \textcolor{black}{0.068} \\
AgACI & \metric{0.923}{0.00} & \textcolor{black}{\metric{71.731}{0.49}} & \textcolor{black}{\metric{67.475}{0.46}} & \textcolor{black}{\metric{0.941}{0.00}} & \textcolor{black}{\secondbest{\metric{0.836}{0.01}}} & \textcolor{black}{\best{\metric{0.896}{0.01}}} & \textcolor{black}{\secondbest{\metric{0.892}{0.01}}} & \textcolor{black}{\secondbest{0.031}} \\
DtACI & \textcolor{black}{\best{\metric{0.902}{0.00}}} & \textcolor{black}{\metric{65.670}{0.37}} & \textcolor{black}{\metric{64.542}{0.55}} & \metric{0.922}{0.00} & \textcolor{black}{\metric{0.786}{0.02}} & \textcolor{black}{\thirdbest{\metric{0.879}{0.01}}} & \textcolor{black}{\metric{0.872}{0.01}} & \textcolor{black}{0.046} \\
SF-OGD & \metric{0.137}{0.00} & \metric{4.243}{0.00} & \textcolor{black}{\metric{4.244}{0.00}} & \textcolor{black}{\metric{0.167}{0.01}} & \textcolor{black}{\metric{0.065}{0.02}} & \textcolor{black}{\metric{0.055}{0.02}} & \metric{0.077}{0.01} & \textcolor{black}{0.809} \\
SCP & \metric{0.923}{0.00} & \textcolor{black}{\metric{73.708}{0.26}} & \textcolor{black}{\metric{73.656}{0.27}} & \textcolor{black}{\metric{0.950}{0.00}} & \textcolor{black}{\metric{0.791}{0.01}} & \textcolor{black}{\metric{0.878}{0.01}} & \textcolor{black}{\metric{0.879}{0.00}} & \textcolor{black}{0.050} \\
PI & \textcolor{black}{\secondbest{\metric{0.898}{0.00}}} & \textcolor{black}{\metric{67.745}{0.83}} & \textcolor{black}{\thirdbest{\metric{64.181}{1.25}}} & \thirdbest{\metric{0.918}{0.00}} & \textcolor{black}{\metric{0.825}{0.02}} & \textcolor{black}{\metric{0.858}{0.02}} & \textcolor{black}{\metric{0.859}{0.01}} & \textcolor{black}{0.044} \\
PID & \textcolor{black}{\thirdbest{\metric{0.898}{0.00}}} & \textcolor{black}{\metric{68.244}{0.91}} & \textcolor{black}{\metric{64.357}{1.15}} & \textcolor{black}{\metric{0.918}{0.00}} & \textcolor{black}{\metric{0.833}{0.02}} & \textcolor{black}{\metric{0.853}{0.02}} & \textcolor{black}{\metric{0.857}{0.01}} & \textcolor{black}{0.044} \\
EnbPI & \metric{0.911}{0.00} & \textcolor{black}{\secondbest{\metric{64.492}{0.65}}} & \textcolor{black}{\metric{64.505}{0.65}} & \metric{0.939}{0.00} & \textcolor{black}{\metric{0.728}{0.01}} & \textcolor{black}{\metric{0.823}{0.01}} & \textcolor{black}{\thirdbest{\metric{0.887}{0.00}}} & \textcolor{black}{0.075} \\
SPCI & \textcolor{black}{\metric{0.746}{0.01}} & \textcolor{black}{\metric{40.973}{0.45}} & \textcolor{black}{\metric{37.432}{0.54}} & \textcolor{black}{\metric{0.773}{0.01}} & \textcolor{black}{\metric{0.645}{0.03}} & \textcolor{black}{\metric{0.582}{0.04}} & \textcolor{black}{\metric{0.726}{0.02}} & \textcolor{black}{0.219} \\
\midrule
\multicolumn{9}{@{}l}{\textit{Capital Bikeshare}}\\*
\textcolor{black}{TRACK} & \textcolor{black}{\metric{0.893}{0.00}} & \textcolor{black}{\best{\metric{1.120}{0.00}}} & \textcolor{black}{\best{\metric{0.978}{0.01}}} & \textcolor{black}{\secondbest{\metric{0.884}{0.00}}} & \best{\metric{0.896}{0.01}} & \textcolor{black}{\secondbest{\metric{0.935}{0.01}}} & \textcolor{black}{\thirdbest{\metric{0.923}{0.01}}} & \textcolor{black}{\secondbest{0.020}} \\
\textcolor{black}{\mbox{TRACK-Abs}} & \metric{0.898}{0.00} & \secondbest{\metric{1.258}{0.00}} & \secondbest{\metric{1.092}{0.01}} & \textcolor{black}{\best{\metric{0.887}{0.00}}} & \textcolor{black}{\secondbest{\metric{0.910}{0.00}}} & \textcolor{black}{\best{\metric{0.884}{0.02}}} & \textcolor{black}{\best{\metric{0.904}{0.01}}} & \textcolor{black}{\best{0.013}} \\
ACI & \textcolor{black}{\thirdbest{\metric{0.899}{0.00}}} & \textcolor{black}{\metric{1.416}{0.01}} & \textcolor{black}{\metric{1.401}{0.02}} & \textcolor{black}{\metric{0.801}{0.00}} & \textcolor{black}{\metric{0.983}{0.00}} & \textcolor{black}{\metric{1.000}{0.00}} & \textcolor{black}{\metric{0.992}{0.00}} & \textcolor{black}{0.094} \\
AgACI & \metric{0.907}{0.00} & \textcolor{black}{\metric{1.444}{0.01}} & \textcolor{black}{\metric{1.402}{0.01}} & \textcolor{black}{\metric{0.818}{0.00}} & \textcolor{black}{\metric{0.983}{0.00}} & \textcolor{black}{\metric{1.000}{0.00}} & \textcolor{black}{\metric{0.992}{0.00}} & \textcolor{black}{0.089} \\
DtACI & \textcolor{black}{\metric{0.897}{0.00}} & \textcolor{black}{\metric{1.391}{0.01}} & \textcolor{black}{\metric{1.395}{0.01}} & \textcolor{black}{\metric{0.797}{0.00}} & \textcolor{black}{\metric{0.982}{0.00}} & \textcolor{black}{\metric{1.000}{0.00}} & \textcolor{black}{\metric{0.992}{0.00}} & \textcolor{black}{0.094} \\
SF-OGD & \metric{0.873}{0.00} & \textcolor{black}{\metric{1.269}{0.00}} & \textcolor{black}{\metric{1.270}{0.00}} & \textcolor{black}{\metric{0.751}{0.00}} & \textcolor{black}{\thirdbest{\metric{0.976}{0.00}}} & \textcolor{black}{\metric{0.999}{0.00}} & \textcolor{black}{\metric{0.991}{0.00}} & \textcolor{black}{0.104} \\
SCP & \textcolor{black}{\metric{0.903}{0.00}} & \metric{1.432}{0.01} & \metric{1.432}{0.01} & \textcolor{black}{\metric{0.807}{0.00}} & \textcolor{black}{\metric{0.985}{0.00}} & \textcolor{black}{\metric{1.000}{0.00}} & \textcolor{black}{\metric{0.992}{0.00}} & \textcolor{black}{0.093} \\
PI & \secondbest{\metric{0.900}{0.00}} & \textcolor{black}{\metric{1.515}{0.02}} & \textcolor{black}{\metric{1.523}{0.02}} & \textcolor{black}{\metric{0.806}{0.00}} & \textcolor{black}{\metric{0.979}{0.00}} & \textcolor{black}{\metric{0.999}{0.00}} & \textcolor{black}{\metric{0.987}{0.01}} & \textcolor{black}{0.090} \\
PID & \best{\metric{0.900}{0.00}} & \textcolor{black}{\metric{1.512}{0.02}} & \textcolor{black}{\metric{1.520}{0.02}} & \textcolor{black}{\metric{0.807}{0.00}} & \textcolor{black}{\metric{0.979}{0.00}} & \textcolor{black}{\metric{0.998}{0.01}} & \textcolor{black}{\metric{0.987}{0.01}} & \textcolor{black}{0.089} \\
EnbPI & \metric{0.911}{0.00} & \textcolor{black}{\thirdbest{\metric{1.308}{0.03}}} & \textcolor{black}{\thirdbest{\metric{1.308}{0.03}}} & \textcolor{black}{\thirdbest{\metric{0.827}{0.01}}} & \textcolor{black}{\metric{0.983}{0.00}} & \textcolor{black}{\thirdbest{\metric{0.985}{0.03}}} & \textcolor{black}{\metric{0.998}{0.00}} & \textcolor{black}{\thirdbest{0.085}} \\
SPCI & \textcolor{black}{\metric{0.738}{0.01}} & \metric{0.818}{0.01} & \textcolor{black}{\metric{0.771}{0.01}} & \textcolor{black}{\metric{0.641}{0.01}} & \textcolor{black}{\metric{0.821}{0.01}} & \textcolor{black}{\metric{0.752}{0.04}} & \textcolor{black}{\secondbest{\metric{0.884}{0.03}}} & \textcolor{black}{0.128} \\
\bottomrule
\end{tabular}}
\caption{\textcolor{black}{One-step-ahead prediction: real data.}}
\label{tab:real-one-selected}
\end{table}

{\color{black}\hypersetup{allcolors=black,citecolor=blue}%
Most methods attain marginal coverage close to nominal in both
applications. \textcolor{black}{However, this overall performance can conceal substantial
differences in coverage across covariate regions.} The signed residual method \textcolor{black}{TRACK}
and the absolute residual method \textcolor{black}{\mbox{TRACK-Abs}} both offer competitive
coverage and interval widths. \textcolor{black}{For both applications, \mbox{TRACK-Abs} provides more
balanced regional coverage, while TRACK produces shorter intervals.} \textcolor{black}{By adapting to local
uncertainty, \textcolor{black}{\textcolor{black}{TRACK} and \mbox{\textcolor{black}{TRACK-Abs}}} reduce the regional imbalances
that marginal coverage alone fails to reveal.}
}%

\subsubsection{\textcolor{black}{Multi-step-ahead prediction}}\label{sec:real-multi-step}
Table~{\color{black}\ref{tab:real-multi-selected}} extends the evaluation to
simultaneous prediction at $H=3$ and $H=10$. We examine whether the
regional calibration gains persist over longer forecast horizons
and whether they are accompanied by smaller prediction sets.

\begin{table}[!htbp]
\centering\setstretch{1}\fontsize{9}{10.5}\selectfont
\setlength{\tabcolsep}{1.5pt}
\renewcommand{\arraystretch}{1.03}
\fitnumtable{\begin{tabular}{@{}l*{8}{c}@{}}
\toprule
Method & Global & Mean & Median & Region~1 & Region~2 & Region~3 & Region~4 & R-MACE \\
\midrule
\multicolumn{9}{@{}l}{\textit{Hackberry wind, $H=3$}}\\*
\textcolor{black}{\mbox{TRACK-J}} & \textcolor{black}{\best{\metric{0.900}{0.00}}} & \textcolor{black}{\best{\metric{127.053}{0.24}}} & \textcolor{black}{\best{\metric{116.707}{0.42}}} & \best{\metric{0.910}{0.00}} & \textcolor{black}{\best{\metric{0.889}{0.01}}} & \textcolor{black}{\best{\metric{0.878}{0.01}}} & \textcolor{black}{\best{\metric{0.877}{0.01}}} & \textcolor{black}{\best{0.017}} \\
Conformal-MS & \metric{0.965}{0.00} & \textcolor{black}{\metric{152.243}{0.37}} & \textcolor{black}{\metric{152.469}{0.45}} & \textcolor{black}{\metric{0.976}{0.00}} & \textcolor{black}{\metric{0.911}{0.01}} & \metric{0.964}{0.01} & \textcolor{black}{\metric{0.942}{0.00}} & 0.048 \\
\midrule
\multicolumn{9}{@{}l}{\textit{Hackberry wind, $H=10$}}\\*
\textcolor{black}{\mbox{TRACK-J}} & \best{\metric{0.904}{0.00}} & \textcolor{black}{\best{\metric{204.874}{0.22}}} & \textcolor{black}{\best{\metric{192.504}{0.49}}} & \textcolor{black}{\best{\metric{0.917}{0.00}}} & \textcolor{black}{\best{\metric{0.843}{0.01}}} & \textcolor{black}{\best{\metric{0.875}{0.01}}} & \best{\metric{0.883}{0.00}} & \best{0.029} \\
Conformal-MS & \metric{0.978}{0.00} & \textcolor{black}{\metric{234.875}{0.28}} & \textcolor{black}{\metric{235.373}{0.31}} & \metric{0.982}{0.00} & \metric{0.967}{0.01} & \metric{0.982}{0.00} & \metric{0.963}{0.00} & \textcolor{black}{0.073} \\
\midrule
\multicolumn{9}{@{}l}{\textit{Capital Bikeshare, $H=3$}}\\*
\textcolor{black}{\mbox{TRACK-J}} & \textcolor{black}{\best{\metric{0.902}{0.00}}} & \textcolor{black}{\best{\metric{2.353}{0.01}}} & \textcolor{black}{\best{\metric{2.054}{0.01}}} & \textcolor{black}{\best{\metric{0.894}{0.00}}} & \textcolor{black}{\best{\metric{0.906}{0.00}}} & \textcolor{black}{\best{\metric{0.911}{0.01}}} & \textcolor{black}{\best{\metric{0.924}{0.01}}} & \textcolor{black}{\best{0.012}} \\
Conformal-MS & \metric{0.923}{0.00} & \metric{2.645}{0.01} & \metric{2.645}{0.01} & \textcolor{black}{\metric{0.863}{0.00}} & \textcolor{black}{\metric{0.971}{0.00}} & \textcolor{black}{\metric{0.989}{0.01}} & \textcolor{black}{\metric{1.000}{0.00}} & \textcolor{black}{0.075} \\
\midrule
\multicolumn{9}{@{}l}{\textit{Capital Bikeshare, $H=10$}}\\*
\textcolor{black}{\mbox{TRACK-J}} & \best{\metric{0.895}{0.00}} & \textcolor{black}{\best{\metric{3.925}{0.01}}} & \textcolor{black}{\best{\metric{3.475}{0.01}}} & \textcolor{black}{\best{\metric{0.884}{0.00}}} & \textcolor{black}{\best{\metric{0.902}{0.00}}} & \textcolor{black}{\best{\metric{0.970}{0.01}}} & \textcolor{black}{\best{\metric{0.892}{0.00}}} & \textcolor{black}{\best{0.024}} \\
Conformal-MS & \metric{0.908}{0.00} & \textcolor{black}{\metric{3.988}{0.01}} & \textcolor{black}{\metric{3.992}{0.01}} & \textcolor{black}{\metric{0.924}{0.00}} & \textcolor{black}{\metric{0.890}{0.00}} & \textcolor{black}{\metric{1.000}{0.00}} & \textcolor{black}{\metric{0.876}{0.01}} & \textcolor{black}{0.040} \\
\bottomrule
\end{tabular}}
\caption{\textcolor{black}{Multi-step-ahead prediction: real data.}}
\label{tab:real-multi-selected}
\end{table}

{\color{black}\hypersetup{allcolors=black,citecolor=blue}%
\textcolor{black}{Across both datasets, \textcolor{black}{\mbox{TRACK-J}} attains marginal joint coverage close to
nominal at $H=3$ and $H=10$, with lower R-MACE and smaller mean and
median root volumes than Conformal-MS. \textcolor{black}{Conformal-MS is generally
conservative at the marginal level, while its regional coverage can vary
appreciably. This distinction highlights the importance of assessing
how joint coverage is distributed across covariate regions. \textcolor{black}{The smaller
mean and median root volumes for \textcolor{black}{\mbox{TRACK-J}} indicate efficiency gains
in both average and typical set size.}} These results show that local
calibration can improve regional coverage in simultaneous prediction
while using smaller prediction sets.}%
}%

\subsection{\textcolor{black}{Summary}}\label{sec:numerical-summary}
{\color{black}\hypersetup{allcolors=black,citecolor=blue}%
{\color{black}\hypersetup{allcolors=black,citecolor=blue}%
Overall, most methods perform well in the homoscedastic benchmark.
The advantages of local calibration become more evident when variance
depends on covariates or the prediction problem involves asymmetric
errors and nonlinear temporal structure. TRACK and its variants adapt
to variation in the local error distribution, improving regional
coverage while often reducing interval width or \textcolor{black}{prediction volume}.
These benefits extend to real data and longer forecast horizons, where
satisfactory marginal coverage alone can mask regional imbalances.
}%
}%

\endgroup

\section{Discussion}\label{sec:disc}

{\color{black}\hypersetup{allcolors=black,citecolor=blue}%
This work addresses the problem of obtaining conditionally valid predictions
from a single time series as predictor memory and forecast horizons grow.
To our knowledge, our analysis gives the first sharp minimax characterization
of \textcolor{black}{score calibration} over \textcolor{black}{given} scoring rules with a prescribed window
span. The matching upper and lower bounds quantify the calibration cost of
score overlap in the worst case. The results for individual rules further
show that this cost depends on how the observations are used, even when the
window span is unchanged. For one-step and joint multi-step forecasting, \textcolor{black}{TRACK and its variants}
provide explicit control of \textcolor{black}{calibration error and coverage discrepancy}
at finite sample sizes. As the sample size increases, coverage approaches
the nominal level conditional on \textcolor{black}{the new covariate value}, with the multi-step
guarantee applying to the forecast trajectory \textcolor{black}{over the specified forecast horizon}.
{\color{black}\hypersetup{allcolors=black,citecolor=blue}These bounds extend to
more general predictors, including those fitted to all past observations,
provided that they admit an accurate rolling-window approximation.}
\textcolor{black}{Experiments show} improved regional coverage
with competitive prediction set sizes.

\color{black}%
Further work could extend the \textcolor{black}{TRACK} framework to nonstationary time series by
combining localization in time and covariates to track changes in the
conditional score distribution. A key question is how
distributional drift and temporal dependence jointly affect coverage. \textcolor{black}{Second, with
high-dimensional covariates, sparsity or low-dimensional structure may aid
calibration, provided that dimension reduction retains information needed
for conditional coverage and the analysis accounts for the error in learning
that structure.} We leave these extensions for future research.
\par%
}%

\Needspace{6\baselineskip}
\renewcommand\refname{References}
\bibliographystyle{apalike}
\begingroup\color{black}\hypersetup{allcolors=black,citecolor=black}%
\setlength{\bibsep}{7pt plus 1pt minus 1pt}
\bibliography{Refs}
\endgroup%
\end{document}